\documentclass[a4paper,11pt]{article}
\pdfoutput=1
\usepackage{jcappub}
\usepackage{feynmp-auto,expdlist}
\usepackage{amsmath, amsfonts, amssymb}
\usepackage{graphicx}
\usepackage{enumerate}
\usepackage{hyperref}
\usepackage{latexsym}
\usepackage{dsfont}
\usepackage{hepnicenames}
\usepackage{enumerate}
\usepackage{soul}
\usepackage[normalem]{ulem}
\usepackage{bbold}
\usepackage{wasysym}
\usepackage{makecell}
\usepackage{multirow}
\usepackage{placeins}
\usepackage{lscape}
\usepackage{pdflscape}

\usepackage{mathrsfs}
\usepackage{braket}
\usepackage{amsmath}
\usepackage{slashed}
\usepackage{amssymb}
\usepackage{epsfig}
\usepackage{graphicx}
\usepackage{color}
\usepackage{rotating}
\usepackage{hyperref}
\usepackage[table,xcdraw,dvipsnames]{xcolor}

\newcommand{\X}{{\cal X}}

\newcommand{\gae}{g_{ae}}

\newcommand{\mX}{\mathcal{X}}

\def\eq#1{{Eq.~(\ref{#1})}}
\def\eqs#1#2{{Eqs.~(\ref{#1})--(\ref{#2})}}
\def\fig#1{{Fig.~\ref{#1}}}

\def\Table#1{{Table~\ref{#1}}}

\def\sect#1{{Sec.~\ref{#1}}}

\def\app#1{{App.~\ref{#1}}}

\def\vev#1{\left\langle #1\right\rangle}

\def\det{\mbox{det}\,}

\renewcommand{\bar}{\overline}

\newcommand{\beq}{\begin{equation}}
\newcommand{\eeq}{\end{equation}}
\newcommand{\bea}{\begin{eqnarray}}
\newcommand{\eea}{\end{eqnarray}}

\renewcommand{\[}{\left[}
\renewcommand{\]}{\right]}
\renewcommand{\(}{\left(}
\renewcommand{\)}{\right)}

\def \gsim{\mathrel{\vcenter
     {\hbox{$>$}\nointerlineskip\hbox{$\sim$}}}}

\title{Stellar Evolution confronts Axion Models}

\author{Luca Di Luzio$^{a,b,c}$,} 
\author{Marco Fedele$^d$,} 
\author{Maurizio Giannotti$^e$,} 
\author{Federico Mescia$^f$,}
\author{Enrico Nardi$^g$}

\affiliation{$^a$ Dipartimento di Fisica e Astronomia `G.~Galilei', Universit\`a di Padova, Italy}
\affiliation{$^b$ Istituto Nazionale Fisica Nucleare, Sezione di Padova, Italy}
\affiliation{$^c$ DESY, Notkestra\ss e 85, D-22607 Hamburg, Germany}
\affiliation{$^d$ Institut f\"ur Theoretische Teilchenphysik, Karlsruhe Institute of Technology, D-76131 Karlsruhe, Germany}
\affiliation{$^e$ Physical Sciences, Barry University, 11300 NE 2nd Ave., Miami Shores, FL 33161, USA}
\affiliation{$^f$ Departament de F\'isica Qu\`antica i Astrof\'isica, Institut de Ci\`encies del Cosmos (ICCUB), Universitat de Barcelona, Mart\'i i Franqu\`es 1, E-08028 Barcelona, Spain}
\affiliation{$^g$ INFN, Laboratori Nazionali di Frascati, C.P. 13, 100044 Frascati, Italy}

\emailAdd{luca.diluzio@unipd.it}
\emailAdd{marco.fedele@kit.edu}
\emailAdd{mgiannotti@barry.edu}
\emailAdd{mescia@ub.edu}
\emailAdd{enrico.nardi@lnf.infn.it}

\abstract{Axion production from astrophysical bodies is a topic in 
continuous development, because of theoretical progress in the  estimate of stellar emission rates
and, especially, because of improved stellar observations.
We carry out a comprehensive analysis of the most informative 
astrophysics data, revisiting  the  bounds on axion couplings to photons, 
nucleons and electrons, and  reassessing the significance of 
various hints of anomalous stellar energy losses. 
We confront the performance of various theoretical constructions in 
accounting for these hints, while complying 
with the observational limits on  axion couplings.  
We identify the most favorable models, and  the regions in the  mass/couplings  parameter space  which are preferred by the global fit. 
Finally, we scrutinize the discovery potential for such models at upcoming helioscopes, namely IAXO and its scaled versions.}

\begin{document} 
\maketitle

\section{Introduction}
\label{sec:intro}

The most elegant solution to the strong CP problem, the so called Peccei-Quinn (PQ) mechanism~\cite{Peccei:1977hh,Peccei:1977ur}, implies the existence of the axion~\cite{Weinberg:1977ma,Wilczek:1977pj}, 
which also provides one of the best particle physics candidates 
for cold dark matter (DM)~\cite{Abbott:1982af,Dine:1982ah,Preskill:1982cy}. 
The axion arises as the pseudo Nambu-Goldstone boson
of a spontaneously broken global Abelian 
symmetry endowed with a mixed anomaly with the color gauge group $SU(3)_c$,  
and the couplings and the mass of the axion are inversely proportional to the 
scale at which spontaneous symmetry breaking 
occurs. Already shortly after 
the axion hypothesis was conceived it became clear that this scale had to lie 
much beyond the electroweak scale, given that no axion signatures could be detected 
in laboratory experiments,\footnote{For  a  historical  account  of  early  experimental 
axion searches  see  e.g.~Sec.~3  in  Ref.~\cite{Davier:1986ps}.} thus implying 
that axions must be very weakly coupled to ordinary matter and very light. 
Such light particles could then be produced in the hot and dense plasma of star cores
and then freely escape, thus providing an additional channel for energy losses from astrophysical 
bodies that would affect their evolution. In fact, the strongest limits on axion couplings 
to electrons, nucleons and electromagnetic radiation come from the requirement that 
stellar lifetimes and energy-loss rates should not conflict with observations~\cite{Raffelt:1996wa,
Giannotti:2017hny, Saikawa:2019lng}.  Astrophysical limits have been derived from the non-observation of axion emission 
from our Sun, from the concordance between the prediction of stellar evolution codes 
and direct  observations of star distributions in the color-magnitude diagram (CMD)
for evolved low-mass star populations, such as red giant branch (RGB) and  horizontal-branch (HB) 
stars in globular clusters, from limits on cooling rates of  white dwarfs (WD) and neutron 
stars (NS),  and from the duration of the neutrino burst from the collapsed core of supernova (SN) 
SN 1987A.
These limits have been frequently revised and updated in the literature, with the coming of better  
observations and more detailed analyses.
However, intriguingly, 
rather than pushing the bounds on the various couplings to smaller and smaller values, 
the more recent astrophysical observations have  hinted at finite, though small, axion couplings. 
More specifically, a set of astrophysical \emph{anomalies}
seem to indicate a preference for non-vanishing couplings of axions to electrons and, to a lesser extent, to photons.
Although individually the significance of each one of these hints is marginal 
(1 or $2\,\sigma$), there is concordance among all the independent observations 
when interpreted in terms of axions~\cite{Giannotti:2017hny}, 
which raises the overall significance at about the $3\,\sigma$ level.
Analyses of other \emph{new physics} candidates does not show the same level of 
agreement among the different observations~\cite{Giannotti:2015dwa}, 
making axions the most interesting candidates to explain the anomalies. 

 In the last few years, the advent of new observations of remarkably improved 
 accuracy has required the revision of some astrophysical bounds on axions. 
In particular, the RGB bound on the axion-electron coupling has been considerably improved and the  significance of the hint has been reduced~\cite{Straniero:2020iyi,Capozzi:2020cbu}.
The supernova and neutron star bounds have also been updated in the last few years~\cite{Carenza:2019pxu,Hamaguchi:2018oqw,Beznogov:2018fda,Sedrakian:2018kdm,Leinson:2021ety}. 
In light of these recent progresses, it seems timely to revise the global analysis 
of the stellar bounds/hints on axions, and to provide updated regions in  
parameter space where the experimental search may be particularly motivated.
This is the task we address in this paper. 

Besides a general, model independent analysis of the stellar observables, we carry out dedicated analyses focusing on specific axion models. Going beyond the approach followed in Ref.~\cite{Giannotti:2017hny}, the present study is not limited  
 to scrutinize only the canonical DFSZ models, but also considers a set of non-universal 
 realizations of DFSZ models, wherein same-type quarks of different generations can couple 
 to different Higgs doublets~\cite{DiLuzio:2017ogq,Bjorkeroth:2018ipq,Bjorkeroth:2019jtx}. In particular, we focus on models that feature the property of nucleo-phobia, i.e.~the capability to strongly suppress the axion coupling to nucleons. 
 Given the tight constrains stemming from observations of SN and NS, which strongly bound the 
 axion-nucleon coupling, these models perform particularly well in reproducing the 
 global set of astrophysics data. We will constrain the parameter region and ascertain the quality of the fit for each of the axion models of our representative sample, singling out those constructions  
that most successfully account for the observations.

As we shall show, the parameter space preferred by stars is the meV mass region (roughly, 1 to 100 meV).
Intriguingly, this parameter region is the goal of the next generation of axion helioscopes. 
In particular, BabyIAXO, expected to be operative in the next few years, may already be able to dig
into the parameter space hinted by stars for some QCD axion models. 
These regions will be better accessible by the full scale IAXO helioscope. 
The even larger IAXO+ will be able to essentially cover the entire parameter space preferred by stars for most well motivated axion models.

The paper is organized as follows. In Sec.~\ref{sec:axionEFT} we review the axion effective Lagrangian and we define the coupling of the axion to the SM particles. 
In Sec.~\ref{sec:astroobs} we give an updated summary of the astrophysical observables relevant to our analysis. A review of the models scrutinized in this paper is given in Sec.~\ref{sec:models}. 
Our results are reported in Sec.~\ref{sec:fits}, with Sec.~\ref{sec:cooling fits} devoted to a 
discussion of the constrains on the axion couplings to SM fields in the various models, and 
Sec.~\ref{sec:haloscopes}  dedicated to analyse, for each model,  the discovery potential
at forthcoming axion search experiments. 
In Sec.~\ref{sec:concl} we resume our results and draw the conclusions.
A detailed calculation of the axion-photon couplings for the non-universal DFSZ models  
is presented in  Appendix~\ref{sec:generalDFSZ}.

\section{Axion effective Lagrangian}
\label{sec:axionEFT}

In this section we focus on the most relevant 
axion couplings from the point of view of astrophysics 
and experimental sensitivities for an axion mass scale ranging 
in the several meV region. 
The axion effective Lagrangian including photons and 
matter fields $f=p,n,e$ 
(defined at a scale $\Lambda < m_p$)
can be written as 
\begin{equation} 
\label{eq:Laint1}
\mathcal{L}_a = \frac{1}{2} (\partial_\mu a)^2 -\frac{1}{2} m_a^2 a^2
+ \frac{\alpha}{8 \pi} \frac{C_{a\gamma}}{f_a} a F_{\mu\nu} \tilde F^{\mu\nu}
+ C_{af} \frac{\partial_\mu a}{2 f_a} \bar f \gamma^\mu \gamma_5 f 
\, ,
\end{equation}
where the adimensional coefficients $C_{a\gamma}$ and $C_{af}$ read \cite{diCortona:2015ldu,DiLuzio:2020wdo}
\begin{align}
\label{eq:Cagamma}
C_{a\gamma} &= \frac{E}{N} - 1.92(4) \, , \\
\label{eq:Cap}
C_{ap} &= -0.47(3) + 0.88(3) \, C_u - 0.39(2) \, C_d - C_{a,\,  {sea}}
\, , \\
\label{eq:Can}
C_{an} &= -0.02(3) + 0.88(3) \, C_d - 0.39(2) \, C_u - C_{a,\,  {sea}}
\, , \\ 
\label{eq:Casea}
C_{a,\,  {sea}} &= 0.038(5) \, C_s 
+0.012(5) \, C_c + 0.009(2) \, C_b + 0.0035(4) \, C_t \, , \\
\label{eq:Cae}
C_{ae} &= C_e + \frac{3\alpha^2}{4\pi^2} 
\[ \frac{E}{N} \log\( \frac{f_a}{m_e} \)
- 1.92(4)
\log\( \frac{ {\rm GeV}}{m_e} \) \] \, . 
\end{align}
In Eqs.~\eqref{eq:Cagamma} and~\eqref{eq:Cae}
$E$ and $N$ correspond  respectively to the 
electromagnetic and QCD anomaly coefficients 
of the PQ current
\beq 
\label{eq:dJPQAnomal}
\partial^\mu J^{\rm PQ}_{\mu} = 
\frac{\alpha_s N}{4\pi} 
G^a_{\mu\nu} \tilde G^{a\, \mu\nu}
+\frac{\alpha E}{4\pi} 
F_{\mu\nu} \tilde F^{\mu\nu} 
\, .  
\eeq
The  coefficients $C_{\psi}$ (with $\psi$ a SM quark or lepton) 
can be written as 
\beq 
\label{eq:Cmix}
 C_\psi = c^0_\psi + \epsilon_\psi \,,
\eeq
where $c^0_\psi$, which is defined in terms of the derivative interaction terms
\beq 
\label{eq:delaaxialq}
c^0_{\psi}\,
\frac{\partial_\mu a}{2 f_a} 
\bar \psi \gamma^\mu \gamma_5 \psi \, ,
\eeq
denotes the  UV axion-fermion couplings 
as determined from the model-dependent PQ charges, 
while  $\epsilon_\psi$ is a correction due to 
possible fermion mixing effects that will generally arise, after electroweak symmetry breaking,  
in models in which the PQ charges are 
generation dependent. In all the models we will consider these corrections 
are small in the case of the quarks of the first generation. In some cases they can become sizeable 
for quarks of the second and third generation (see Table~\ref{tab:axionmodels}) 
but since their contribution as sea quarks to $C_{ap},\,C_{an}$ is already suppressed by small coefficients 
(see \eq{eq:Casea}) we can expected that \eqs{eq:Cap}{eq:Can}, which are derived under the assumption of 
no mixing effects, will still remain valid to a good approximation. Finally, 
 in Eq.~\eqref{eq:Cae} the first term $C_e=c^0_e+\epsilon_e$ corresponds to the tree level axion 
 coupling to the electron, 
while the second term is the one-loop contribution originating from a triangle loop involving two photons 
\cite{Srednicki:1985xd,Chang:1993gm}.

In the following, 
we will use the following rescaled axion couplings
\beq 
\label{eq:gagammagaf}
g_{a\gamma} = \frac{\alpha}{2 \pi} \frac{C_{a\gamma}}{f_a} \, , \qquad 
g_{af} = C_{af} \frac{m_f}{f_a} 
\, , 
\eeq
and employ the definitions 
$g_{\gamma 10} \equiv g_{a\gamma} \times \left(10^{10} \,  \text{GeV}\right)$ 
and 
$g_{e 13} \equiv g_{ae} \times 10^{13}$. 
Finally the relation between the axion mass and the axion decay constant 
reads \cite{Gorghetto:2018ocs} 
\beq 
\label{eq:axionmassfa}
m_a = 5.691(51) \( \frac{10^{12} \ \text{GeV}}{f_a} \) \, \text{$\mu$eV} \, . 
\eeq

\section{Astrophysical observables and cooling anomalies}
\label{sec:astroobs}

Astrophysical considerations have played a quite significant role
in the investigation of the physics of light, weakly interacting particles, such as neutrinos,  dark photons, axions, axion like particles (ALPs) etc.~\cite{Raffelt:1996wa}.   
Constraints derived from the observations of stellar populations  
are often much tighter than bounds from direct searches. 
Intriguingly,  a series of anomalous astrophysical observations have led
to speculations that new physics may be at play in determining the details of  stellar evolution~\cite{Giannotti:2015dwa,Giannotti:2015kwo,Giannotti:2017hny,Hoof:2018ieb,DiVecchia:2019ejf,DiLuzio:2020wdo,Agrawal:2021dbo}. 
The axion (or ALP) case is especially compelling
since, contrarily to other new physics candidates,
fits particularly well 
all  astrophysical observations~\cite{Giannotti:2015kwo,Giannotti:2017hny}. 
A list of updated bounds and hints from stellar evolution on the axion
couplings are summarized in Table~\ref{tab:Astro_bounds}. 
The aim of this Section is to review and update the status of the  impact of axion 
emission on stellar evolution. 
We begin by reviewing the most relevant axion production mechanisms in stars, 
and next we summarize the constraints and hints from observations of different 
stellar systems. 

\begin{table}[t]
	\setlength\tabcolsep{4pt}
	\begin{center}
		\begin{tabular}{ l c c c}
			Star  				&  Hint 										&  Bound  					& Reference	 		\\ \hline
			\ Sun \phantom{$\Big|$} &   No Hint 												    &  $g_{\gamma10}\leq 2.7$ 	&	\cite{Vinyoles:2015aba}	\\
			\vspace{0.2cm}
			WDLF 				&  $g_{e13}=1.5^{+0.3}_{-0.5}$ 					&  $g_{e13}\leq 2.1$ 	&	\cite{Bertolami:2014wua}  	\\
			\vspace{0.2cm}
			WDV 		 	 	&  $g_{e13}=2.9^{+0.6}_{-0.9}$ 		 			&  $g_{e13}\leq 4.1$ 	 &	 \cite{Corsico:2019nmr}	 \\
			\vspace{0.2cm}
			RGBT	(22 GGCs)	&  $g_{e13}=0.60^{+0.32}_{-0.58}$ 	            &  $g_{e13}\leq 1.5$        &     \cite{Straniero:2020iyi}    \\	 
			\vspace{0.2cm}
			RGBT	(NGC 4258)  &  No Hint 		         				                        &  $g_{e13}\leq 1.6$ 	&	\cite{Capozzi:2020cbu}  \\
			\vspace{0.2cm}
			HB 					&  $g_{\gamma10}=0.3^{+0.2}_{-0.2}$ 	&  $g_{\gamma10}\leq 0.65$ 	&	\cite{Ayala:2014pea,Straniero:2015nvc}\\
			\vspace{0.2cm}
			SN 1987A 			&  No Hint 	&  $ g_{aN}\lesssim 9.1 \!\times\! 10^{-10}  $ 		&	 \cite{Carenza:2019pxu}\\ 
			\vspace{0.2cm}
			NS (CAS A) 			&  No Hint 	&  $ (g_{ap}^2+1.6\, g_{an}^2)^{1/2}\lesssim 1.0\!\times\! 10^{-9} $ 		&	\cite{Hamaguchi:2018oqw} \\ 
			\vspace{0.2cm}
						NS (CAS A) 			&  No Hint 	&  $g_{an}\lesssim 3\!\times\! 10^{-10}$		&	\cite{Leinson:2021ety} \\ 
			\vspace{0.2cm}
			NS (HESS) 	&  No Hint 	&  $g_{an}\leq 2.8\!\times\! 10^{-10}$		&	\cite{Beznogov:2018fda} \\ 
			\hline
		\end{tabular} 
\caption{
Updated bounds and hints from stellar evolution on the axion couplings. 
	WDLF: White Dwarf Luminosity Function; 
	WDV: White Dwarf Variables; 
	RGBT: Red Giant Branch Tip;
	HB: Horizontal Branch;
	SN: Supernova; NS: Neutron Star.
	The bound from SN 1987A is on the effective coupling 
	$g_{aN}$ defined in~\eq{eq:gan_gap_SN_bound}. 
	The two RGBT bounds are from two independent analyses, the first one based on 22 galactic globular clusters~\cite{Straniero:2020iyi} and the second based on the NGC 4258 galaxy~\cite{Capozzi:2020cbu}. 
    The three NS bounds are from independent analyses, 
	two based on observations of CAS~A~\cite{Hamaguchi:2018oqw,Leinson:2021ety} 
	and one on observations of HESS J1731-347~	\cite{Beznogov:2018fda}. We have not listed the hint from the CAS~A study in Ref.~\cite{Leinson:2014ioa} since it is in tension with the more recent bounds shown in the Table.
	In the last column we give the references to the original analyses.}
	\label{tab:Astro_bounds}
	\end{center}
\end{table}

\subsection{Axion production in stars}
\label{sec:production}

The most relevant axion production mechanisms in stellar environments are the Primakoff, Compton, and electron bremsstrahlung processes. A pedagogical introduction to these processes can be found in Ref.~\cite{Raffelt:1996wa}, while a set of approximate expressions and numerical results can be 
found in the appendix of Ref.~\cite{Straniero:2019dtm}. Here we just recall a few semi-analytical  
formulae  for axion production through the different processes that will be useful for the following 
discussion. 

In star core plasma with a relatively low density the most relevant axion production mechanisms 
are the Primakoff and Compton processes. 
The former is induced by the axion coupling to photons and the latter by its coupling to electrons. 
The Primakoff process is the  photon conversion into an axion in the electric field of  electrons or 
ions in the plasma: 
\begin{align}
\gamma+ Ze\to a+ Ze \,.
\end{align}
Neglecting degeneracy effects and the plasma frequency (a good assumption in  
environments in which  the Primakoff process is the dominating axion production mechanism) 
it is possible to provide a semi-analytical expression for the energy-loss rate per unit mass due  to axion emission~\cite{Friedland:2012hj}:
\begin{align}
\label{eq:Primakoff_approx}
\varepsilon_P\simeq 2.8\times 10^{-31} F(\xi) \left( \frac{g_{a\gamma}}{\rm GeV^{-1}} \right)^{2}
\frac{T^7}{\rho}\, {\rm erg\,g^{-1}\,s^{-1}}\,,
\end{align}
where $T$ and $\rho$ are in K and in g cm$^{-3}$.
The function $F$ depends on the Debye-Huckel screening wavenumber
$\kappa_{S}$ via the variable $\xi\equiv\kappa_{S}/2T$, and  
can be explicitly  expressed as an integral over the photon distribution (see Eq.~(4.79) in Ref.~\cite{Raffelt:1990yz}).
An  approximate expression, which agrees with numerical results at better than 2\% over the entire 
range of $\xi$, is~\cite{Friedland:2012hj} 
\begin{equation}
\label{eq:alternativefit}
F(\xi)\simeq 
\left(\frac{1.037\,\xi^{2}}{1.01+0.185\,\xi^{2}}+\frac{1.037\,\xi^{2}}{44+0.628\,\xi^{2}}\right)
\ln\left(3.85+\frac{3.99}{\xi^{2}}\right)\,.
\end{equation}
In general, $F(\xi)$ is ${\cal O}(1)$ for relevant stellar conditions. 
More general numerical recipes, valid also in a degenerate medium, can be found in the appendix of Ref.~\cite{Straniero:2019dtm}.

In general the Primakoff process does not play a significant 
role in the evolution of superdense star cores, in which case other processes dominate.  The Compton process 
\begin{align}
\gamma +e\to  \gamma +a
\end{align}  
accounts for the production of axions from the scattering of thermal photons on electrons.
The Compton axion emission rate is a steep function of the temperature:
\begin{align}
\varepsilon_{\rm C}\simeq 2.7\times 10^{-22} g_{ae}^2 \frac{1}{\mu_e}\left( \frac{n_{e}^{\rm eff}}{n_e} \right)\,T^6 \,{\rm erg\,g^{-1}\,s^{-1}} \,,
\end{align}
where $\mu_e=\left( \sum X_j Z_j/A_j \right)^{-1}$ is the mean molecular weight per electron with $ X_j $  the relative mass density of the $j$-th ion and  $ Z_j,~A_j $ its charge and mass number respectively,   
 $n_e$ is the number density of electrons while $n_{e}^{\rm eff}$ is the effective number density of electron targets.
At high densities, degeneracy effects reduce $n_{e}^{\rm eff}$, suppressing the Compton rate.
Thus, this process is particularly effective in high-temperature environments, as long as the 
density is still relatively low to prevent electron degeneracy (cf.~Fig.~1 in~\cite{DiLuzio:2020jjp}).
At higher densities and especially  when electron degeneracy conditions are reached, the  most efficient 
axion production mechanism is the electron/ion bremsstrahlung process
\begin{align}
e +Ze\to  e + Ze +a\,.
\end{align}  
For degenerate plasma conditions, the axion energy-loss rates per unit mass 
can be approximated as  
\begin{align}
\varepsilon_{\rm B}
\simeq 8.6 \times 10^{-7} F_B\, g_{ae}^{2} T^{4}\left(\sum\frac{X_j Z_j^2}{A_j}\right)  {\rm erg\,g^{-1}\,s^{-1}}\,.
\label{eq_bremsstrahlung_degenerate}
\end{align}
The mild density dependence of the degenerate rate is accounted for by the dimensionless function $F_B$.
An explicit expression for this function can be found in~\cite{Raffelt:1994ry} (see also Sec.~3.5 of Ref.~\cite{Raffelt:1996wa} for a pedagogical presentation).
Numerically its value is  of order one for the 
typical stellar plasma conditions in which 
the degenerate bremsstrahlung process dominates $\rho \sim 10^{5}-10^{6}\, {\rm g\,cm^{-3}}$ and $T\sim 10^{7}-10^{8}\,{\rm K}$.

In a nuclear medium such that of a SN
collapsed core or in NS cores, 
the processes discussed above are generally sub-dominant~\cite{Lucente:2021hbp}. 
If the axion coupling to nucleons is not particularly suppressed, as is the case in most axion models,  
a more efficient production mechanism is axion bremsstrahlung in    
nucleon-nucleon  collisions
\begin{align}
& N+N^{\prime}\to N+N^{\prime}+a\,,
\end{align}
where $N,N^{\prime}=n,p$.
If we model the nucleon-nucleon interaction with the exchange of a single pion, 
it is evident that the pion mass in the propagator will suppress the emission rate, unless the temperature is such that the typical momentum  exchanged in the collision, which is of the order of the nucleon momentum $q_N\sim (3m_NT)^{1/2}$, is larger than the pion mass. 
This demands $T\gtrsim 10$ MeV, a temperature typical of SN and NS  cores.
Therefore, among the various stellar objects,  SN and NS 
provide the best environments to test the axion nucleon coupling.

Approximate expressions for the $nn$ scattering emission rates (the $pp$ scattering is similar) in the limit of non-degenerate (ND) and degenerate (D) nuclei are given below~\cite{Raffelt:1996wa}
\begin{subequations}\label{eq_nuclear_bremsstrahlung_approx}
\begin{align}
& \varepsilon_{{\rm ND}}\approx 2.0\times 10^{38} g_{an}^2 \rho_{14}\,T_{30}^{3.5}\, {\rm erg}\, {\rm g}^{-1} {\rm s}^{-1} \,,\\
& \varepsilon_{{\rm D}}\approx 4.7\times 10^{39} g_{an}^2 \rho_{14}^{-2/3}\,T_{30}^{6}\, {\rm erg}\, {\rm g}^{-1} {\rm s}^{-1} \,,
\end{align}
\end{subequations}
where $T_{30}=T/30$MeV and $\rho_{14}=\rho/10^{14}{\rm g} \, {\rm cm}^{-3}$.
These expression are based on a series of approximations,\footnote{The equations were derived assuming 
that the nucleon-nucleon interactions are described through the one-pion-exchange potential, and assume that pions are massless.
Furthermore, they assume that the plasma is either fully degenerate or completely non-degenerate.
All these assumptions were relaxed 
in Ref.~\cite{Carenza:2019pxu}.} 
however, they still provide a good order of magnitude estimate of the axion emission rate from a SN or NS, and 
in particular they put in evidence the steeper temperature dependence in the case of a degenerate medium. 
However, throughout  this work we use the more accurate 
numerical results from Ref.~\cite{Carenza:2019pxu} 
(see Sec.~\ref{sec:SN}).

\subsection{Axion Bounds from White Dwarfs}
\label{sec:WDs}

WDs represent the last stage of the evolution of low mass stars,
following the exhaustion of the nuclear fuel. 
Therefore, during this phase the star is just cooling.
WDs are characterized by a dense core of degenerate electrons, 
with typical density of about
$10^6$ g$\,$cm$^{-3}$ and a core temperature dependent on the age of the star. 
Young WDs are hot and brighter, and cool through volume neutrino emission. 
At later times, the photon surface cooling dominates.

The addition of exotic particles, such as axions, can have a strong impact on the evolution of the WD, accelerating the stellar cooling. 
Thus, testing the cooling efficiency is an indirect way to probe the existence of new physics. 
There are two independent ways to test the cooling efficiency of WDs.
One is to observe the WD luminosity function (WDLF), which shows the WD number distribution in different luminosity bins (see Sec.~\ref{sec:WDLF}).
The other, is to measure changes in the period of WD variables (WDV), a class of WD whose luminosity periodically changes with time (see Sec.~\ref{sec:WDV}).

Both methods indicate a preference for some unidentified cooling, which could well be provided by  axion emission provided  their coupling to electrons is of a few $10^{-13}$. Below 
we give more details on the analyses of WD cooling rates.

\subsubsection{White Dwarf luminosity function}
\label{sec:WDLF}

As we have already mentioned, the WDLF represents the number distribution of WDs as a function of their luminosity. 
This distribution has been used for decades as a tool to measure 
the WD cooling efficiency, since the number of WDs with a certain luminosity obviously depends 
on how efficiently the star looses energy. 
If axions exist, they would be produced in a WD core primarily through the bremsstrahlung process.  
Hence, the WDLF has provided information about the axion coupling to electrons (see Ref.~\cite{Isern:2020non} for a recent review).
Unfortunately, observations from the Sloan Digital Sky Survey (SDSS) and the and SuperCOSMOS Sky Survey (SSD), on which the current studies are based, are not consistent within their
quoted error bars, indicating that the systematic uncertainties in these observations might have been
underestimated~\cite{MillerBertolami:2014oki}. 
Consequently, there
 is no complete consensus on the exact  bound on the axion-electron coupling 
derived from observations of the WDLF. 
A summary of different studies can be found in Ref.~\cite{Bertolami:2014wua}.
In that analysis, the authors 
enlarged the error bars to take into account not only their systematic uncertainties, but also the discrepancies between the SDSS and SSS observations (see discussion therein, in sec. 4.2).
In the present study, we adopt the result 
$g_{ae}\leq 2.1\times 10^{-13}$ at $(2\,\sigma)$, derived in Ref.~\cite{Bertolami:2014wua}.

Furthermore, most studies of the WDLF seem to indicate an excessive cooling with respect
to the standard prediction. 
Such cooling has been interpreted as due to an axion-electron coupling
$ g_{ae}\simeq (1.4\pm 0.3) \times 10^{-13} $ (at 1$ \sigma $)~\cite{Bertolami:2014wua}.\footnote{Such result is not free from controversy. 
A later study of the hot part of the WDLF~\cite{Hansen:2015lqa} did not confirm this anomalous behavior.
However, the hotter section of the WDLF has much larger observational errors and the 
axion (or ALP) production would be almost completely hidden by standard neutrino 
cooling in the hottest WDs. 
The most recent work~\cite{Isern:2018uce} seems to confirm an excessively efficient cooling, which can be explained by the emission of axions.  Alternatively, it has been recently proposed 
that the excess of cooling could be explained by  neutrino pair synchrotron emission enhanced 
by extremely large magnetic fields, in excess 
of $10^{11}$ Gauss, confined in the WD cores~\cite{Drewes:2021fjx}. This possibility can in principle be bounded by dedicated asteroseismology surveys.}

\subsubsection{White Dwarf Variables}
\label{sec:WDV}

The WDV are a set of WDs whose luminosity changes periodically. 
The period $P$ ranges from a few to several minutes, depending on the particular star.
It is well known that observations of the secular change, $\dot{P}$, of the WDV period
provide information about the efficiency of the WD cooling. 
In fact, to a very good approximation
$\dot P/P$ is directly proportional to the cooling rate  $\dot T/T$. 
Hence, an accurate measurement of  $\dot{P}$
allows to set bounds on possible sources of extra cooling  
(see Ref.~\cite{Corsico:2019nmr} for a comprehensive review).

For over two decades, observations of the period decrease 
($ \dot P /P $) of particular WDVs
have shown discrepancies with the expected behavior.
This is clear from the data in Table~\ref{tab:Pdot}, which shows the results
of the observations of the WDVs analyzed so far~\cite{Corsico:2019nmr}. 
The systematic tendency of the observed $\dot P$ to be larger than the expected values
can be interpreted in terms of a new particle, produced in the core and efficiently 
carrying energy outside the star.
Even though specific observations have been interpreted in various ways, 
for example in terms of an anomalously large neutrino magnetic moment (see, e.g., Ref.~\cite{Corsico:2014mpa}),
a global analysis of all the data indicates a preference for axions among other WISPy candidates~\cite{Giannotti:2015kwo},
and identifies the coupling with electrons in the range $g_{ae}=2.9^{+0.6}_{-0.9}\times 10^{-13}$ (at 1$\sigma$)~\cite{Giannotti:2017hny}, with a 2$\sigma$ bound of $g_{ae}\leq 4.1\times 10^{-13}$, as reported in Table~\ref{tab:Astro_bounds}.
The analysis in this paper is based on the recent review~\cite{Corsico:2019nmr}, where the most updated studies on the viable WDV are considered. 
The results for the axion couplings are summarized in Tab.~\ref{tab:Pdot}.

We conclude observing that, following the approach usually employed in the literature and first outlined in Ref.~\cite{Giannotti:2015kwo}, we will not include in our fits data relative to the WD
G117-B15A: the analysis for this star shares many theoretical similarities to the one relative to R548, but the experimental results are somewhat different, with the hint on $g_{ae}$ stemming from the former noticeably stronger than the one inferred from the latter. Given that results relative to these two WDs are based on very similar hypothesis, we will therefore conservatively include in our fit only data pertaining to R548, along with the data from PG 1351+489 and L 19-2.
\begin{table}[t]
	\begin{center}
			 \setlength\tabcolsep{4pt}
		\begin{tabular}{ l c  c c c  c c}
			Star  				&  $ P $(s)&  $ \dot{P}_{\rm obs} $(s/s) 		&  $ \dot{P}_{\rm th} $(s/s) 			& $g_{ae}^{(\rm best)}$   	&  $g_{ae}^{\rm (max)}(2\sigma)$ \\ \hline
			G117 - B15A\phantom{$^{\displaystyle |}$}			&  215 	& $ (5.5 \pm 0.8)\!\times\! 10^{-15} $	&  $ (1.25 \pm 0.09)\!\times\! 10^{-15} $		&  $5.6\!\times\! 10^{-13}$ 	& $ 6.7\!\times\! 10^{-13} $  \\
			R548 		 	 	&  213 	&  $ (3.3 \pm 1.1)\!\times\! 10^{-15} $ 	&  $ (1.1 \pm 0.09)\!\times\! 10^{-15} $ 		&  $4.8 \!\times\! 10^{-13} $ 	&  $6.8 \!\times\! 10^{-13} $ \\
			PG 1351+489			&  489 	& $ (2.0 \pm 0.9)\!\times\! 10^{-13} $ 	& $ (0.81 \pm 0.5)\!\times\! 10^{-13} $ 		& $ 2.1 \!\times\! 10^{-13} $ 	& $ 3.8 \!\times\! 10^{-13} $ \\
			L 19-2	(113) 		&  113 	&  $ (3.0\pm 0.6)\!\times\! 10^{-15} $ 	&  $ (1.42 \pm 0.85)\!\times\! 10^{-15} $		&  $5.1 \!\times\! 10^{-13} $ 	&  $7.7 \!\times\! 10^{-13} $\\
			L 19-2 	(192) 		&  192 	&  $ (3.0\pm 0.6)\!\times\! 10^{-15} $ 	&  $ (2.41\pm 1.45) \!\times\! 10^{-15} $		&  $ 2.5 \!\times\! 10^{-13} $ 	&  $6.1  \times 10^{-13} $ \\ 
			\hline
		\end{tabular}
		\caption{
		Measured and expected value of $\dot P$ for a set of WDV.
		Here, $P$ is the period of the variable star and $\dot P$ its time derivative.
		The interpretation in terms of axion coupling to electrons is also shown. 	
		(Data from Ref.~\cite{DiLuzio:2020wdo} except for G117-B15A, which corresponds to the updated analysis in Ref.~\cite{Kepler_2020}).}
		\label{tab:Pdot}
	\end{center}
\end{table}

\subsection{Axion Bounds from the tip of RGB stars in globular cluster}
\label{sec:tipRGB}

After completing the evolution through the main sequence, characterized by a H-burning core, 
low mass stars begin climbing the RGB, 
well visible in the color-magnitude diagram as a diagonal line starting at the main sequence
and directed toward the colder and higher luminosity region.
During the evolution in the RGB, stars are characterized by a He core and a burning H shell,
whose ashes increase the He core mass, while the star luminosity (determined by
equilibrium at the surface of the 
He core between  thermal pressure supporting the non-degenerate envelope
against the gravity pull from the core) keeps growing.
The process continues until the core reaches sufficiently large 
temperatures and  densities 
($T \sim 10^8\,$K, $\rho=10^6\,$g\,cm$^{-3}$)
to ignite He, an event known as the He-flash. 
At this stage the star has reached the maximum luminosity,  
that is the RGB tip (RGBT), after which it shrinks and moves to the HB. 
If an additional core-cooling mechanism is at play,  He ignition is delayed, 
the core would accrete a larger mass, and the star would reach 
higher luminosities. Therefore, measurements of the luminosity of the RGB tip
allow to test the rate of cooling during the RGB phase.  
The method is particularly effective for constraining $\gae$ 
since in red giant cores axions can be efficiently produced 
via electron bremsstrahlung.

We denote by $M_{I,{\rm TRGB}}$ 
the luminosity of the tip of the RGB  in globular clusters (GC). 
Following Ref.~\cite{DiLuzio:2020jjp}, based on the analysis in Ref.~\cite{Viaux:2013lha,Giannotti:2015kwo}, 
it is possible to derive the following analytical expression for the expected magnitude of the RGBT:
\beq
\label{eq:MITRGB_theo}
M_{I,{\rm TRGB}}^{\rm theo} = - 4.08 
- 0.25 \bigg(\sqrt{g_{e13}^2 + 0.96^2} 
- 0.96 - 0.17 g_{e13}^{1.5}\bigg) \,, 
\eeq
that has an associated theoretical uncertainty
$\sigma^2 ={0.039^2 + (0.046 + 0.012 g_{e13})^2}$.
This should be compared with the observational values. 
The latest analyses are those in Refs.~\cite{Straniero:2020iyi,Capozzi:2020cbu}.
The first, is based on the global analysis of a sample of 22 GC while the second 
analyzes individual clusters and galaxies. 
Both studies give very similar results (cf.~Table~\ref{tab:Astro_bounds}). 
Here, as a reference, we consider the observational value
found in the comprehensive analysis from Ref.~\cite{Straniero:2020iyi}.
This choice results in the bound on the axion-electron coupling $g_{e13}\leq 1.5$ (2$\,\sigma$),
which is perfectly compatible with the one found in Ref.~\cite{Capozzi:2020cbu},
but also hints to a non-vanishing value for  $g_{e13}$. 
At any rate, we have verified that removing this additional hint from our fit does not alter 
significantly the results.

\subsection{Helium burning stars}
\label{sec:Rparam}

After  He ignition, the RG core expands and the star migrates to the 
horizontal branch 
of the color magnitude diagram, characterized by a He burning non-degenerate core. 
The core of a HB star has a density of about $\rho\sim 10^4\, {\rm g \,cm^{-3}}$, which is about two orders of magnitude less than that of a RGB star or a WD. 
In this environment, axions are efficiently produced through the Primakoff and Compton processes. 
The effect of the additional energy loss provided by  axion production and 
emission is to accelerate the helium consumption in the HB core and,
consequently, to reduce the lifetime of this stage. 
A very efficient way to probe this effect is by measuring the so called $R$ parameter, 
$R=N_{\rm HB}/N_{\rm RGB}$, which measures the ratio 
between the number in the HB and in the upper portion of the RGB 
in GC. 

Historically, measurements of the $R$ parameter have been used to derive bounds on the axion-photon
coupling~\cite{Raffelt:1985nk,Raffelt:1987yu,Ayala:2014pea,Straniero:2015nvc}.
However, the axion electron coupling can also impact the $R$ parameter. 
Following the results of Refs.~\cite{Giannotti:2015kwo,Giannotti:2017hny,DiLuzio:2020wdo}, 
we present the expected $R$ parameter in the following form 
\begin{align}
	\label{eq:R-parameter}
	R=R_0(Y)-
	F_{a\gamma}\! \left( g_{\gamma10} \right) -
	F_{ae}\! \left( g_{e13} \right)\,,
\end{align}
where $R_0(Y)$ is a function of the helium abundance ($Y$) in the GC 
and $F_{a\gamma}$, $F_{ae}$ are some positive-definite functions 
of the axion couplings.
Evidently, the impact of axions on the $R$ parameter may be induced by its couplings with
photons as well as by its coupling with electrons.
However, surprisingly, there are no explicit stellar numerical evaluation of the $R$ parameter which include both couplings. 
An approximate expression for the functions $F$, based on a series of older numerical results, was presented in Refs.~\cite{Giannotti:2015kwo,Giannotti:2017hny}. 
There, it was found 
\begin{align}
	\label{eq:R-parameter_functions}
	&R_0(Y)=0.02+7.33\, Y\,,\\
	&F_{a\gamma}(x)=0.095 \sqrt{21.86+21.08 \,x}\,,  \\
	&F_{ae}(x)=0.0053\,x^2+0.039
	\left( \sqrt{1.23^2+x^2} -1.23-0.14\,x^{3/2}\right) . 
\end{align}
The theoretical expression in Eq.~\eqref{eq:R-parameter} should be compared with observational results, in order to constraint the axion couplings. 
Ref.~\cite{Ayala:2014pea},  reported the value 
$R = 1.39 \pm 0.03$ 
from the analysis of 39 clusters, and used the result to derive the bound
$ g_{a\gamma}\leq 0.65\times 10^{-10}\,{\rm GeV^{-1}}$ at $2\,\sigma$~\cite{Ayala:2014pea,Straniero:2015nvc}, under the assumption that the 
axion couples only to photons. 

More massive He burning stars, with mass $ M\sim 8-12 M_{\odot} $, can also provide some insight into the axion-photon coupling~\cite{Friedland:2012hj,Carosi:2013rla}.
The (core) He burning stage of these stars is characterized by a migration towards the blue (hotter) region of the CMD and back.
This journey is known as the \textit{blue loop}. 
The existence of the loop is corroborated by many astronomical observations. 
In particular, this stage is essential to account for the observed Cepheid stars (see, e.g., Ref.~\cite{Kippenhahn:1994wva}).
Ref.~\cite{Friedland:2012hj}, based on numerical simulations of 
solar metallicity stars in the $ 8-12 M_{\odot} $ mass range, 
showed that a coupling larger than  
$ \approx 0.8\times 10^{-10} {\rm GeV}^{-1} $ 
would cause the complete disappearance of the blue loop 
while a somewhat lower values of $g_{a\gamma} $ might help explain 
the observed deficiency of blue with respect to red supergiants, as is discussed, e.g., in Ref.~\cite{McQuinn:2011bb}.
The numerous uncertainties in the microphysics and in the numerical description of the blue loop stage, however, have not permitted a more quantitative assessment of this possibility~\cite{Giannotti:2015dwa}.
Hence, these results will be ignored in our present work.

\subsection{Supernovae}
\label{sec:SN}

The observation of the neutrino signal from SN 1987A, 
and the recent observation of a NS associated with it~\cite{Page:2020gsx}
supported the picture of a neutrino driven SN explosion with  neutrinos carrying away 
about 99\% of the  energy released in the explosion.
Since new weakly coupled particles could accelerate the cooling and reduce the observed duration of the  neutrino signal, SN 1987A nutrino data have been widely used to constrain models of new physics,
and particularly axion properties~\cite{Turner:1987by,Burrows:1988ah,Raffelt:1987yt,Raffelt:1990yz}. 

The plasma in the core of a SN, in the first few seconds after the explosion, is extremely hot ($T\sim 30$ MeV) and dense ($\rho\sim 3\times 10^{14} \,$g/cm$^3$).
This makes the SN an extremely interesting environment to study new physics. 
Axions can be produced in this environment through different processes, including the Primakoff process~\cite{Brockway:1996yr,Payez:2014xsa,Grifols:1996id}, and the electron bremsstrahlung~\cite{Calibbi:2020jvd}.
However, for couplings allowed by other astrophysical considerations
the nuclear bremsstrahlung, driven by the axion coupling to neutrons and protons, 
largely dominates over the latter two.\footnote{It was recently pointed out 
in Ref.~\cite{Carenza:2020cis,Fischer:2021jfm} that pion processes 
$\pi^- + p \to a+n$ might dominate the axion production rate, 
becoming even more efficient than the nuclear bremsstrahlung 
at some temperatures and densities.
This process has been discussed for a long time in the literature~\cite{Turner:1991ax,Raffelt:1993ix,Keil:1996ju}.
 Only recently, however, it was shown that  pion abundance in the early stages of a SN is much larger than 
previously expected~\cite{Fore:2019wib}, perhaps to the point of making the pion production mechanism
the dominant one~\cite{Carenza:2020cis}.
We have not included these results in our analysis since 
the current studies, although focused on the observational consequences of the pion-induced axion production mechanism, do not derive new bounds on the axion-nucleon couplings~\cite{Fischer:2021jfm}.
}
The most recent analysis of the axion-nucleon bremsstrahlung~\cite{Carenza:2019pxu} give the bound
\begin{equation} 
g^2_{aN} \equiv g_{an}^2+ 0.61\, g_{ap}^2 + 0.53\, g_{an}\,g_{ap}\lesssim 8.26 \times 10^{-19} \,,
\label{eq:gan_gap_SN_bound}
\end{equation}
which is the one we adopt in the present study.
If the axion coupling to nucleons is large enough, axions may be trapped 
in the SN core~\cite{Burrows:1990pk,Carenza:2019pxu}.
In this case, the cooling efficiency is reduced and the axion bound weakened. 
Unfortunately, the exact value of the coupling for which axions are sufficiently trapped so that  
the observed duration of the neutrino signal is not reduced, is afflicted by several uncertainties. 
However, in most models in which the SM
fermions carry a PQ charge the trapping
regime is already excluded by other stellar bounds. 
Thus, in the present work  we will not consider further this possibility.

\subsection{Neutron stars}
\label{sec:NS}

Observations of the cooling of NS 
also provide information about the axion-nucleon coupling~\cite{Keller:2012yr,Sedrakian:2015krq,Hamaguchi:2018oqw,Beznogov:2018fda,Sedrakian:2018kdm,Leinson:2021ety}. 
Several NS bounds on $g_{aN}$ exist in the literature, 
however, they are not always consistent with one another. 
One of the most studied NS is the one in CAS A.
In recent years, there has been some speculation that its 
anomalously rapid cooling 
could be a hint of axions with coupling to neutrons~\cite{Leinson:2014ioa}
$g_{an}\simeq 4\times 10^{-10}$.
However, the data can also be explained assuming a neutron triplet superfluid transition occurring  at the present time, $t\sim 320$ years, in addition to a proton superconductivity operating at $t\ll 320$ years~\cite{Hamaguchi:2018oqw}.
Under these assumptions, it was possible to fit the available data well, leaving little room for additional axion cooling, corresponding to the bound~\cite{Hamaguchi:2018oqw}: 
\begin{align}
\label{eq:NS_CAS_A}
g_{ap}^2+1.6\, g_{an}^2\leq 1.1\times 10^{-18}\,.
\end{align}
A stronger bound, though only on the axion-neutron coupling, 
\begin{align}
\label{eq:NS_J1731}
g_{an}\leq 2.8\times 10^{-10}\,,
\end{align}
was inferred from observations of the NS in HESS J1731-347~\cite{Beznogov:2018fda}.
This bound is also in good agreement with a newer analysis of
the NS in CAS A carried out in Ref.~\cite{Leinson:2021ety} (cf.~Table~\ref{tab:Astro_bounds}).

In the present work, we take the bound in Eq.\eqref{eq:NS_J1731} as a reference for the 
NS bound on the axion-nucleon coupling.
It should be remarked, however, that 
there is no universal consensus on the NS bound. 
For example,  Ref.~\cite{Sedrakian:2018kdm} recently proposed a considerable less stringent result, 
\begin{align}
\label{eq:NS_Sedrakian}
g_{an}\lesssim (2.5-3.2)\times 10^{-9}\,.
\end{align}
For this reason, we present separate analyses, with and without the NS results.

\subsection{Model-independent fit}
\label{sec:modindepfit}

In this  subsection we describe the global  fit to 
the relevant astrophysical observables described above, 
that we have performed  by taking the axion couplings $g_{ae}$ and $g_{a\gamma}$ 
as uncorrelated parameters.  
Given the absence of any particular assumption about an underling axion model, 
  this first analysis is denoted as {\it model-independent}.
Note that this implies that $g_{ae}$ and $g_{a\gamma}$ are not correlated to 
the couplings to nucleons either, and hence it is consistent to neglect,  
in the model-independent fit, the data inferred from SNe and NS observations.
However, this additional set of data will be taken into account in the model-dependent fits performed in Sec.~\ref{sec:fits} where, given the model, all the couplings are correlated in a well defined way. 
In short, in this subsection we will consider only  data pertaining to WD, RGB and HB stars.
\begin{figure}[!t]
\centering
\includegraphics[width=9cm]{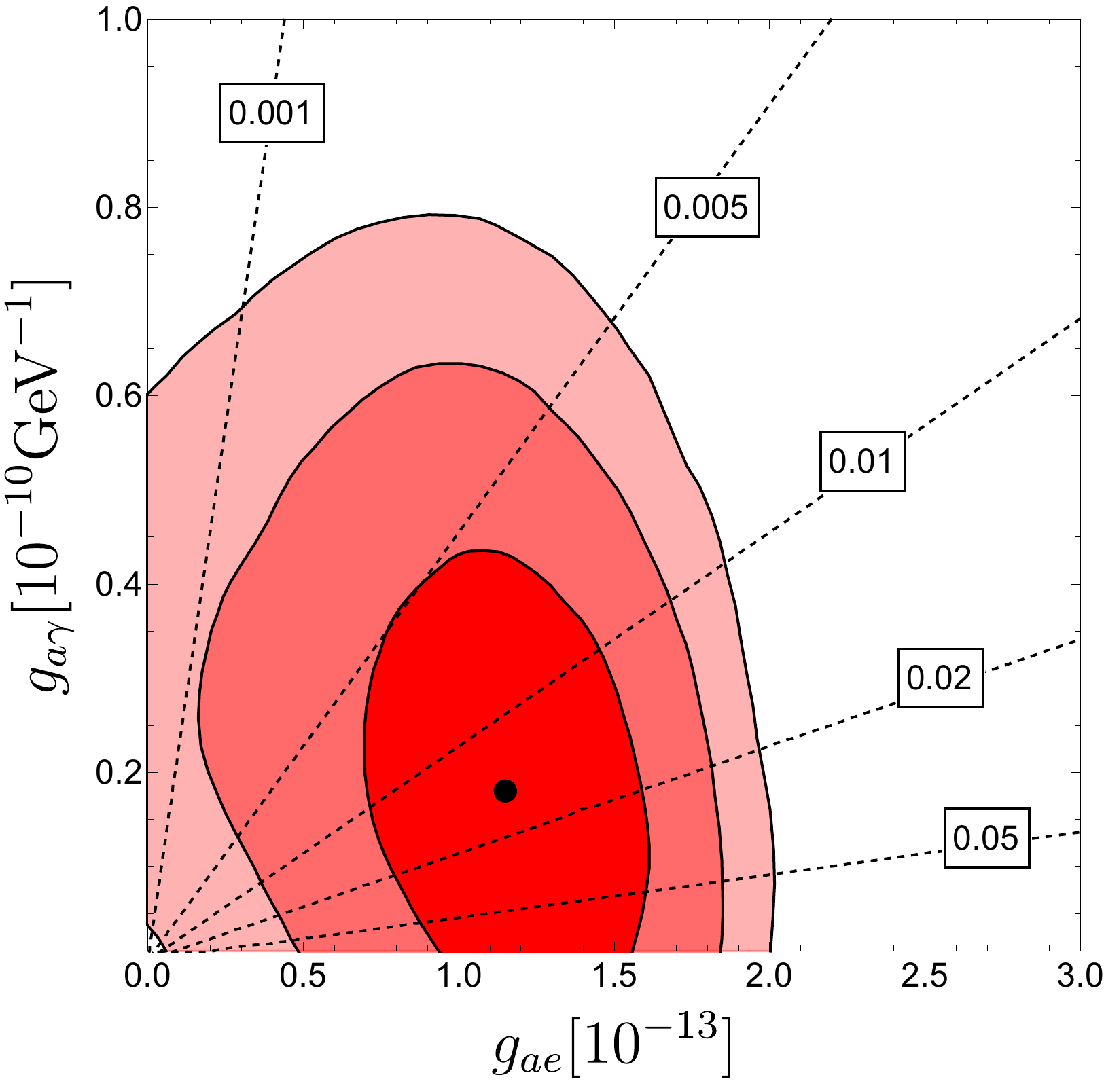}
\caption{Global bounds 
from stellar cooling: $1\sigma$ (dark red) to   
$3\sigma$ (light red). The global mode, corresponding to $g_{e12}=0.1$ and $g_{\gamma10}=0.2$, is shown as a black dot. Dashed lines represent iso-lines
for the ratio $C_{ae}/C_{a\gamma}$.
\label{fig:gaegag}
}
\end{figure}

We depict in Fig.~\ref{fig:gaegag} the $1\sigma$, $2\sigma$ and $3\sigma$ regions allowed by the fit, showing also  the iso-lines for the ratio $C_{ae}/C_{a\gamma}$. 
The best fit point is marked with a black dot, and corresponds to $g_{ae} \simeq 1.2 \times 10^{-13}$ 
and 
$g_{a\gamma} \simeq 0.18 \times 10^{-10} \text{GeV}^{-1}$, with $C_{ae}/C_{a\gamma} \simeq \mathcal{O}(10^{-2})$. 
It is interesting to notice that the present data, including the information relative to HB, can be accommodated by the introduction of a non-vanishing axion-electron coupling only. 
On the other hand, the SM case 
($g_{ae}=g_{a\gamma}=0$)  is excluded by present data at the about the $ 3\sigma$ level. 

Let us add a comment regarding a comparison with the  global analysis in Ref.~\cite{Giannotti:2017hny}. 
The results of this previous study are qualitatively similar to what is shown in our Fig.~\ref{fig:gaegag}. The most relevant quantitative difference  is the RGB contribution to the global fit.
Ref.~\cite{Giannotti:2017hny} was based on the RGB analysis in Ref.~\cite{Viaux:2013lha}, which adopted an incorrect 
screening prescription for the nuclear reaction rates~\cite{Serenelli2017}.
The correction of this numerical issue in Refs.~\cite{Straniero:2020iyi,Capozzi:2020cbu} strengthened substantially the bound on the axion-electron coupling, 
and caused a shift of the entire $g_{ae}$ region in Fig.~\ref{fig:gaegag} to the left. 
As stated in Sec.~\ref{sec:tipRGB}, we employed in our fit the results obtained in Ref.~\cite{Straniero:2020iyi}, but we tested the stability of our findings under this choice performing also a fit where we used the results from Ref.~\cite{Capozzi:2020cbu}. 
We found that the contours of the preferred regions differ only slightly between the two choices, with the SM case always excluded at the $\sim 3\sigma$ level. The results shown in Fig.~\ref{fig:gaegag}  therefore give a reliable representation of the region preferred by the stars.

As anticipated above, in Sec.~\ref{sec:models} we will review a selection of motivated axion models, each of which will imply
specific correlations among the various axion couplings in terms of a few model parameters. 
The impact of the stellar evolution data on the corresponding bounds and the effectiveness 
of those specific models in accommodating the hints for extra energy emission from stars  
are discussed in Sec.~\ref{sec:fits}.

\section{A representative sample of axion models}
\label{sec:models}

In this Section we introduce a set  of explicit axion models
which yield different axion couplings to the nucleons, the electrons and 
the photons. In Sec.~\ref{sec:cooling fits}
we will consider how well they can perform in addressing  the issue of possible 
anomalies in stellar energy losses, accommodating the observational hints  
while respecting all other phenomenological bounds.

The model-independent analysis 
of \sect{sec:modindepfit}  
suggests that promising candidates among axion models should comply with a  first requirement 
of predicting a sizeable $C_{ae} / C_{a\gamma}$ ratio. 
For this reason KSVZ 
models \cite{Kim:1979if,Shifman:1979if}, 
in which the $C_{ae}$ coupling arises 
radiatively from a triangle loop involving two photons, and hence is 
induced by $ C_{a\gamma}$, are not well suited to 
explain the cooling hints. 
Indeed, for KSVZ models one obtains
\beq 
\label{eq:KSVZest}
\frac{C_{ae}}{C_{a\gamma}} \simeq  
\frac{3\alpha^2}{4\pi^2} 
\frac{\frac{E}{N} \log\( \frac{f_a}{m_e} \)
- 1.92
\log\( \frac{ {\rm GeV}}{m_e} \)}
{E/N - 1.92 } 
\simeq 
\frac{3\alpha^2}{4\pi^2} \log\( \frac{f_a}{m_e} \) 
\simeq 1.1 \times 10^{-4}
\, , 
\eeq
where in the next-to-last relation  we have selected 
the log-enhanced contribution and assumed $E/N \gg 1.92 $,
and in the last relation we have chosen as  reference value $f_a = 10^{9}$ GeV.     
As it can be seen from  \fig{fig:gaegag}, the iso-line 
 $C_{ae} / C_{a\gamma} = 10^{-3}$ already misses the 1$\sigma$ region preferred 
 by the global fit, so that the KSVZ value of $10^{-4}$  
cannot provide a good fit to the stellar cooling hints. 
Hence, we will not considered the KSVZ model any further in the present study.

A better starting point 
for addressing the cooling hints is provided by
DFSZ models \cite{Zhitnitsky:1980tq,Dine:1981rt}
since the axion coupling to electrons arises at tree level. 
In particular, the 
nucleo-phobic axion models of 
Refs.~\cite{DiLuzio:2017ogq,Bjorkeroth:2018ipq,Bjorkeroth:2019jtx,Badziak:2021apn} 
allow to relax astrophysical bounds from NS/SN, 
while keeping at the same time a sizeable coupling to 
electrons and photons. 
An additional motivation to consider the latter class of models
consists in the fact that their  generation dependent PQ charge assignment is engineered 
in such a way that  the contributions of two generations to the QCD anomaly factor 
cancel out yielding $N = N_1 + N_2 + N_3 = N_1$. This implies a DW number 
$N_{\rm DW} = 2N_1 = 2$ or 1 (rather than  $N_{\rm DW} = 2N = 6$ or 3 as in canonical DFSZ models) \cite{DiLuzio:2017ogq}. 
The importance of the existence of 
DFSZ-like models with $N_{\rm DW} = 1$  consists
in the fact that they are strongly preferred in post-inflationary PQ-breaking scenarios since  they 
are free from the DW problem. In this case in fact the network of axionic strings coupled to a single 
DW spontaneously decays around the time when the axion acquires a mass. 
In this scenario, 
under the assumption that  the shape of the instantaneous axion emission from string decays 
is IR dominated~\cite{Gorghetto:2020qws,Hiramatsu:2010yu}, 
cosmological considerations 
based on the requirement that the axion relic density will not exceed that of the DM,
together with the requirement that axion interactions will be sufficiently suppressed 
to impede axion thermalization into dark radiation, yield a viable window for the axion mass 
between 0.2 and 100 meV~\cite{Hoof:2021jft}.\footnote{The exact value for the lower limit is subject to systematic uncertainties in the simulations of topological defects. 
If instead the axion emission spectrum  
is not IR dominated,  
the leading contribution to the axion relic density comes from the misalignment mechanism,  
and the lower value for the axion mass drops to $m_a\gsim 25\,\mu $eV~\cite{Buschmann:2019icd,Buschmann:2021sdq}. 
In pre-inflationary PQ breaking 
scenarios $N_{\rm DW}>1$ does not constitute a problem since all topological defects are inflated away.
In that case the axion relic density is generated solely by the misalignment mechanism and, 
while axion masses in the meV range cannot be excluded, if DM is dominantly composed by 
axions then 
limits on iso-curvature fluctuations generated  during inflation 
 constrain $m_a$ to lie 
below the meV scale \cite{Wantz:2009it}.} As 
can be seen from Table~\ref{tab:fit_results} in Sec.~\ref{sec:cooling fits}, this agrees well  
with the mass window  favoured by the cooling 
hints.

In this Section we first review the basic features of the 
above-mentioned DFSZ-like axion models, and next we  generalize further the 
classification of axion models with two Higgs doublets and generation dependent PQ charges
by including novel constructions. 
In particular, we show that  models featuring particularly large coupling to photons 
and electrons can still be nucleo-phobic, a feature that 
allows to optimize  the fit to the cooling anomalies while bringing the corresponding axions  
within the discovery potential of future helioscopes such as (Baby)IAXO. 
In spite of the fact that we analyze a reasonably large number of different constructions, 
we stress that our list is far from representing the entire panorama of axion models. In particular, 
it leaves out models in which the axion coupling to electrons can be enhanced by specific mechanisms 
(see e.g.~\cite{Ballesteros:2016xej,Darme:2020gyx,DiLuzio:2021pxd}). 
Nevertheless, it is unlikely that these alternative possibilities could yield much better 
fits to the hinted anomalies than our representative subset. This is because the best fit 
conditions are realised when $g_{ae}/g_{a\gamma}\approx 0.6 \times 10^{-2}\,$GeV and 
$g_{ap},\,g_{an} \approx 0$, and indeed this  parameter space point is  already approached 
in some of the models we have considered.  

\subsection{Universal DFSZ models}
\label{sec:bench}

The scalar sector of DFSZ models \cite{Zhitnitsky:1980tq,Dine:1981rt}
features a complex scalar SM  
singlet $\Phi \sim (1,1,0)$
where the labels in parenthesis refer to transformation properties under the 
SM gauge group $SU(3)_C\times SU(2)_L\times U(1)_Y$,  
and two Higgs doublets 
$H_u \sim (1,2,1/2)$ and $H_d \sim (1,2,-1/2)$ 
that couple respectively to up- and down-type quarks
in a generation-independent way. 
A scalar operator  $H_u H_d \Phi^{\dag 2}$ 
or $H_u H_d \Phi^{\dag}$ 
(corresponding 
respectively 
to 
$N_{\rm DW} = 6$ or 3)
breaks the  $U(1)_\Phi\times U(1)_{H_u}\times U(1)_{H_d}$
re-phasing symmetry of the scalars to $U(1)_{\rm PQ} \times U(1)_Y$ so that 
a PQ symmetry is 
preserved by the 
scalar potential.
There are two possible variants of the model, depending on whether the lepton sector couples to $H_d$ (DFSZ1) 
or to $\tilde H_u = i \sigma^2 H_u^*$ (DFSZ2). For a review see Sec.~2.7.2 in \cite{DiLuzio:2020wdo}. 

\subsubsection{DFSZ1}
\label{sec:DFSZI}

Leaving Yukawa coupling constants understood, the Yukawa sector contains   
the following  operators 
\beq
\label{eq:DFSZ1}
\bar q_i u_j H_u \, ,\ \:\bar q_i  d_j H_d\, , \ \:\bar \ell_i e_j H_d\, , \eeq
where $i,j = 1,2,3$ are generation indices, $q_i,\, \ell_i$ denote the quarks and leptons  
$SU(2)_L$ doublets and $u_i,\, d_i,\, e_i$ the right-handed (RH)  singlets. 
The corresponding axion coupling coefficients are 
\begin{align}
&\frac{E}{N} =\frac{8}{3} \, , \quad
c^0_{u,c,t} =\frac{c^2_\beta}{3}  \, ,\quad c^0_{d,s,b}=\frac{s^2_\beta}{3}\, ,\quad 
c^0_{e,\mu,\tau}=\frac{s^2_\beta}{3}\, , 
\end{align}
with 
$c_\beta \equiv \cos\beta$, 
$s_\beta \equiv \sin\beta$ and 
$\tan\beta = \vev{H_u} / \vev{H_d}\equiv v_u/v_d$. 
Requiring that the quark Yukawa couplings remain in the perturbativity domain restricts  
the vacuum angle $\beta$ to lie withing the   
interval 
$\tan\beta \in [0.25, 170]$
(the perturbativity limits for the models  considered in this work are reviewed in \sect{sec:pert}). 
Note that for both DFSZ1 and DFSZ2  the 
axion couples to the SM fermions in a 
generation-independent way so that  
there are no corrections to the axion couplings from intergenerational mixing effects, and hence 
$C_\psi =c^0_\psi$, see \eq{eq:Cmix}.

\subsubsection{DFSZ2}
\label{sec:DFSZII}

The Yukawa sector contains the  following  operators 
\beq
\label{eq:DFSZ2}
\bar q_i u_j H_u \, ,\ \:\bar q_i  d_j H_d\, , \ \:\bar \ell_i e_j \tilde H_u\, ,   
\eeq
and the corresponding axion coupling coefficients are 
\begin{align}
&\frac{E}{N} = \frac{2}{3} \, , \quad
c^0_{u,c,t} =\frac{c^2_\beta}{3}\, ,\quad c^0_{d,s,b}=\frac{s^2_\beta}{3}\, ,\quad 
c^0_{e,\mu,\tau}=- \frac{c^2_\beta}{3}\, . 
\end{align}
The vacuum angle can range in the same perturbativity interval than in DFSZ1, that is 
$\tan\beta \in [0.25, 170]$.

\subsection{Non-universal DFSZ models}
\label{sec:nonuDFSZ}

We denote as ``non-universal'' those models that have the same 
scalar content than DFSZ1 and DFSZ2 
(two Higgs doublets and one SM singlet $\Phi$) 
but for which same-type quarks of different generations can couple to 
different Higgs doublets.
Clearly, in this case the labels  $u$ and $d$ for the Higgs doublets loose their meaning, 
and it is more convenient  to employ a notation where 
the two scalars have the same quantum numbers,  
and  denote them as $H_{1,2} \sim (1,2,-1/2)$.
 The vacuum angle is defined as $\tan\beta = \vev{H_2}/\vev{H_1}\equiv v_2/v_1$.
For all these models, the requirement that 
the PQ current is orthogonal to the hypercharge current
(which implies orthogonality between the respective Goldstone bosons)  
fixes the  PQ charges 
of the two Higgs doublets 
as $\mX_1=-s^2_\beta$ and
$\mX_2=c^2_\beta$.

In the following, we review a set of models which 
can feature the property of being  nucleo-phobic, 
namely for a specific choice of the vacuum angle $\beta$ the 
coefficients $C_{p,n}$ can be strongly suppressed 
with respect to their natural DFSZ values. 
A non-trivial result, for which a proof is given in 
\app{sec:generalDFSZ}, is that in a general non-universal 
DFSZ model with two Higgs doublets,  the model dependent ratio  
$E/N$ can only span over the following finite set of values: 
$E/N =
\{ 2, \frac{5}{3}, \frac{4}{3}, \frac{8}{3}, \frac{2}{3}, \frac{11}{3}, -\frac{1}{3}, \frac{14}{3}, -\frac{4}{3}, \frac{20}{3}, -\frac{10}{3} \}$. 
Although larger values of $E/N$ can be achieved by 
introducing an arbitrary number of extra Higgs doublets 
(see e.g.~Refs.~\cite{DiLuzio:2016sbl,DiLuzio:2017pfr,Darme:2020gyx}), 
here we focus only on the case of two  (or three, see below) Higgs doublets. 
In particular, for the two Higgs doublet case we consider
the nucleo-phobic models 
M1, M2, M3 and M4 
introduced in Ref.~\cite{DiLuzio:2017ogq},  
and the
$\mathcal{T}_{2}^{(u,d)}$ 
models of Ref.~\cite{Bjorkeroth:2018ipq}. 
The latter feature $E/N$ values 
corresponding to the last two entries 
in the list above, which result in the  
largest coupling $g_{a\gamma}$. 
For the case of three Higgs doublets  we consider 
the nucleo-phobic 3HDM model of Ref.~\cite{Bjorkeroth:2019jtx},  
where the suppression of the axion-nucleon couplings 
is strongly correlated with the value of the axion coupling to the electrons.

Given that in all these models the PQ charge assignments are generation dependent,
the axion couplings to quarks and leptons will in general receive corrections from 
inter-generational fermion mixing. These  will depend on the particular model and on 
specific assumptions on the mixing matrices for the 
left-handed (LH) and right-handed (RH) states. 
As regards the mixing independent part of  the couplings,  they 
can instead be written in a model independent way in terms of the PQ charges of the SM 
fermions as~\cite{DiLuzio:2020wdo}
\begin{align}
&c^0_{u,\,c,\,t} = \frac{\X_{u_i} - \X_{q_i}}{2N}\, ,\quad  c^0_{d,\,s,\,b} = \frac{\X_{d_i} - \X_{q_i}}{2N} \, , \quad 
c^0_{e,\,\mu,\,\tau} = \frac{\X_{e_i} - \X_{\ell_i}}{2N}\, , 
\end{align}
where 
\begin{align}
&2N = \sum_{i=1}^3 (\X_{u_i} +\X_{d_i}-2 \X_{q_i})\,,
\end{align}
is the QCD anomaly factor
which is  not affected by corrections from fermion mixing.

\subsubsection{Non-universal 2+1 models}
\label{sec:2+1}

The M1, M2, M3 and M4 models are characterized by a 2+1 
structure of the PQ charge assignments, namely two generations 
replicate the same set of PQ charges. Note that as explained in 
Ref.~\cite{DiLuzio:2017ogq} in this case all the entries in the 
up- and down-type quark Yukawa matrices  are allowed and there
are no texture zeros.  We recall below for each model 
the structures of the Yukawa operators and the charge assignments for the different generations.
More details can be found in Ref.~\cite{DiLuzio:2017ogq}.

\subsubsection*{M1}
\label{sec:M1}

The Yukawa sector of the M1 model contains the following operators
\begin{align}
&\bar q_1  u_1 H_1\, ,\ \:\bar q_3  u_3 H_{2}\, , \ \:\bar q_1  u_3 H_{1}\, , \ \:\bar q_3  u_1 H_{2}\, , \nonumber  \\
  \label{eq:m1}
&\bar q_1  d_1 \tilde H_2\, , \ \:\bar q_3  d_3 \tilde H_{1}\, ,\ \:\bar q_1  d_3 \tilde H_{2}\, ,\ \:\bar q_3  d_1 \tilde H_{1}\, , \nonumber \\
&\bar \ell_1  e_1 \tilde H_1\, , \ \:\bar \ell_3  e_3 \tilde H_{2}\, , \ \:\bar \ell_1  e_3 \tilde H_{1}\, ,\ \:\bar \ell_3  e_1 \tilde H_{2}\, .
\end{align}
The PQ charges of the first generation are replicated  for the second generation,
which thus appears in an analogous set of operators with the generation label 1 replaced by 2. 
The PQ charge assignments are
\begin{align}
\X_{q_i} &= (0, 0, 1)\, ,\:
\X_{u_i} = (s_\beta^2, s_\beta^2, s_\beta^2)\, ,\: \X_{d_i} = (c_\beta^2, c_\beta^2,c_\beta^2)\, , \nonumber \\
\X_{\ell_i} &=  -\X_{q_i}\, ,\:\X_{e_i}=-\X_{u_i}\, .
\end{align}
The anomaly coefficients and the mixing independent part of  
the axion couplings are
\begin{align}
&\frac{E}{N} = \frac{2}{3}\, ,\quad \quad  \ 2N=1 \, , \nonumber \\
&c^0_{u,c} =s_\beta^2\, ,\quad 
\; \ \ c^0_t=-c_\beta^2\, , \nonumber \\ 
&c^0_{d,s}=c_\beta^2\, ,\quad 
\; \ \ c^0_b=-s_\beta^2\, , \nonumber \\
&c^0_{e,\mu}=-s_\beta^2\, , \quad 
c^0_\tau=c_\beta^2\, ,
\end{align}
with 
$\tan\beta \in [0.25, 170]$.
Since in all the sectors the charges of the RH fields of the three generations are the same, there 
are no RH mixing corrections. In the LH  sector mixing effects enter because the third generation 
has  different charges from the first two. For the quarks, if we assume that the LH mixing matrix 
has CKM-like numerical entries, mixing effects are small and can be neglected. Accordingly, we will 
simply use $C_{u,c,t,d,s,b} \simeq c^0_{u,c,t,d,s,b}$. In the lepton sector instead there are no reasons to 
expect a particular suppression of  $e_L$-$ \tau_L$ mixing, hence  $C_e = c^0_e + \epsilon_{L}$ 
where the expression for 
$\epsilon_{L} \in [-1,1]$ 
can be found in Ref.~\cite{DiLuzio:2017ogq}. 

\subsubsection*{M2}
\label{sec:M2}

The Yukawa sector of the M2 model contains the following operators 
\begin{align}
&\bar q_1  u_1 H_1\, ,\ \:\bar q_3  u_3 H_{1}\, , \ \:\bar q_1  u_3 H_{1}\, , \ \:\bar q_3  u_1 H_{1}\, , \nonumber \\
  \label{eq:m2}
&\bar q_1  d_1 \tilde H_1\, , \ \:\bar q_3  d_3 \tilde H_{2}\, ,\ \:\bar q_1  d_3 \tilde H_{2}\, ,\ \:\bar q_3  d_1 \tilde H_{1}\, , \nonumber \\
&\bar \ell_1  e_1 \tilde H_1\, , \ \:\bar \ell_3  e_3 \tilde H_{2}\, , \ \:\bar \ell_1  e_3 \tilde H_{1}\, ,\ \:\bar \ell_3  e_1 \tilde H_{2} \, .
\end{align}
The PQ charge assignments are
\begin{align}
\X_{q_i} &= (0, 0, 0)\, ,\:
\X_{u_i} = (s_\beta^2, s_\beta^2, s_\beta^2)\, ,\: \X_{d_i} = (c_\beta^2,-s_\beta^2, -s_\beta^2)\, , \nonumber \\
\X_{\ell_i}&=  -(0,0,1)\, ,\:\X_{e_i}=-(s_\beta^2,s_\beta^2,s_\beta^2)\, .
\end{align}
The anomaly coefficients and the mixing independent part of  
the axion couplings are
\begin{align}
&\frac{E}{N} = \frac{8}{3}\, ,\quad \quad \  2N=1 \, , \nonumber \\
&c^0_u=s_\beta^2\, ,\quad \quad
c^0_{c,t}=s_\beta^2\, , \nonumber \\ 
&c^0_d=c_\beta^2\, ,\quad \quad
c^0_{s,b}=-s_\beta^2\, ,  \nonumber \\
&c^0_{e,\mu}=-s_\beta^2\, , \quad 
c^0_\tau=c_\beta^2\, .
\end{align}
Note that for this case the range for the vacuum angle allowed by Yukawa perturbativity  
is $\tan\beta \in [0.0024, 4.0]$. 
As it can be seen from the charge assignments, there are no mixing effects 
in the LH and  up-type  RH quark sectors (hence $C_{u,c,t} = c^0_{u,c,t}$).  
As regards the RH down-type quarks, although there is no phenomenological 
information on the RH mixings, we will assume that the related effects are 
negligible, and we will approximate   $C_{d,s,b} \simeq c^0_{d,s,b}$ while, 
for the leptons, we adopt the same prescription used for M1. 

\subsubsection*{M3}
\label{sec:M3}

The M3  models is defined by the following set of Yukawa operators:
\begin{align}
&\bar q_1  u_1 H_1\, ,\ \:\bar q_3  u_3 H_{2}\, , \ \:\bar q_1  u_3 H_{2}\, , \ \:\bar q_3  u_1 H_{1}\, , \nonumber \\
  \label{eq:m3}
&\bar q_1  d_1 \tilde H_1\, , \ \:\bar q_3  d_3 \tilde H_{1}\, ,\ \:\bar q_1  d_3 \tilde H_{1}\, ,\ \:\bar q_3  d_1 \tilde H_{1}\, , \nonumber \\
&\bar \ell_1  e_1 \tilde H_1\, , \ \:\bar \ell_3  e_3 \tilde H_{2}\, , \ \:\bar \ell_1  e_3 \tilde H_{1}\, ,\ \:\bar \ell_3  e_1 \tilde H_{2} \, ,
\end{align}
where, as in M1 and M2, the quark charges of the second generation replicate those of the 
first one, while for the leptons they replicate the charges of the third generation. 
Explicitly, the PQ charge assignments are
\begin{align}
\X_{q_i} &= (0, 0, 0)\, ,\:
\X_{u_i} = (-c_\beta^2, s_\beta^2, s_\beta^2),\: \X_{d_i} =- (s_\beta^2, s_\beta^2,s_\beta^2)\, , \nonumber \\
\X_{\ell_i} &=  -(0,1,1)\, ,\:\X_{e_i}=-(s_\beta^2,s_\beta^2,s_\beta^2)\, .
\end{align}
The anomaly coefficients and the mixing independent part of  
the  couplings are
\begin{align}
&\frac{E}{N} =-\frac{4}{3}\, ,\quad\quad 2N=-1\, , \nonumber \\
&c^0_u=c_\beta^2\, ,\quad \quad 
c^0_{c,t}=-s_\beta^2\, , \nonumber \\ 
&c^0_d=s_\beta^2\, ,\quad \quad
c^0_{s,b}=s_\beta^2\, , \nonumber \\
&c^0_{e}=s_\beta^2\, , \quad \quad
c^0_{\mu,\tau}=-c_\beta^2\, ,
\end{align}
with $\tan\beta \in [0.0024, 4.0]$.
We can easily read off the charge assignments the following 
relations:  $C_{d,s,b} = c^0_{d,s,b}$,  $C_{u,c,t} \simeq c^0_{u,c,t}$ (assuming RH up-type 
quark mixings are small) and $C_e = c^0_e + \epsilon_{L}$.

\subsubsection*{M4}
\label{sec:M4}

For the M4 models the Yukawa sector contains the following operators
\begin{align}
&\bar q_1  u_1 H_1,\ \:\bar q_3  u_3 H_{1}, \ \:\bar q_1  u_3 H_{1}, \ \:\bar q_3  u_1 H_{1}, \nonumber \\
  \label{eq:m4}
&\bar q_1  d_1 \tilde H_1, \ \:\bar q_3  d_3 \tilde H_{2},\ \:\bar q_1  d_3 \tilde H_{2},\ \:\bar q_3  d_1 \tilde H_{1}, \nonumber \\
&\bar \ell_1  e_1 \tilde H_1, \ \:\bar \ell_3  e_3 \tilde H_{2}, \ \:\bar \ell_1  e_3 \tilde H_{1},\ \:\bar \ell_3  e_1 \tilde H_{2}\,,
\end{align}
where as in M3 for the second generation the quarks replicate the  charges of the first generation,
while the leptons replicate the charges of the third generation. 
The PQ charge assignments are
\begin{align}
\X_{q_i} &= (0, 0, 0),\:
\X_{u_i} = (s_\beta^2, s_\beta^2, s_\beta^2),\: \X_{d_i} =(c_\beta^2,-s_\beta^2,-s_\beta^2), \nonumber \\
\X_{\ell_i}&=  -(0,1,1),\:\X_{e_i}=-(s_\beta^2,s_\beta^2,s_\beta^2)\,.
\end{align}
The anomaly coefficients and the mixing independent part of  
the  couplings are
\begin{align}
&\frac{E}{N} =14/3,\quad \ 2N=1 \nonumber \\
&c^0_u=s_\beta^2,\quad \quad 
c^0_{c,t}=s_\beta^2, \nonumber \\ 
&c^0_d=c_\beta^2,\quad \quad
c^0_{s,b}=-s_\beta^2, \nonumber \\
&c^0_{e}=-s_\beta^2\, , \quad 
c^0_{\mu,\tau}=c_\beta^2\, ,
\end{align}
with $\tan\beta \in [0.0024, 4.0]$. 
The structure of the charge assignments in generation space implies 
$C_{u,c,t} = c^0_{u,c,t}$,  $C_{d,s,b} \simeq c^0_{d,s,b}$ (assuming RH down-type 
quark mixings are small) and $C_e = c^0_e + \epsilon_{L}$.

\subsubsection{1+1+1 models}
\label{sec:1+1+1}

In Ref.~\cite{Bjorkeroth:2018ipq}
two other models, denoted as  $\mathcal{T}_2^{(u,d)}$,  
were considered, that are  characterized by a 1+1+1 
structure of the PQ charges 
in the quark sector, 
namely all generations 
can have different PQ charges.\footnote{The 
PQ charges in the lepton sector 
are assumed for 
simplicity to be universal. 
However, employing 
a generation dependent 
$U(1)$ symmetry for leptons 
can yield non-trivial 
predictions also for 
neutrino masses and mixings \cite{Bjorkeroth:2019rat}.} They correspond to the class of nucleo-phobic models 
for which the Yukawa matrices have the maximum number 
of texture zeros that still allows to reproduce the
quark masses and CKM mixings. 
Below we recall the main features of these two models.

\subsubsection*{$\mathcal{T}_2^{(u)}$}
\label{sec:M6}

The Yukawa sector of the $\mathcal{T}_2^{(u)}$ model contains the following operators 
\begin{align}
& \bar q_1  u_1 H_1\, ,\ \:   \bar q_2  u_2 H_2\, ,\ \:\bar q_3 u_2 H_1\, ,\ \:     \bar q_3  u_3 H_{2}\, , \nonumber \\
  \label{eq:11u}
& \bar q_1  d_1 \tilde H_2\, , \ \:    \bar q_1  d_2 \tilde H_1\, ,  \ \:
  \bar q_2 d_2  \tilde H_{2}\, , \ \:    \bar q_2  d_3 \tilde H_{1}\, , \ \:    \bar q_3  d_3 \tilde H_{2}\, , \nonumber \\
  & \bar \ell_i e_j \tilde H_{1} \, ,
\end{align}
For the quarks,  out of the eighteen operators allowed by the SM gauge symmetry only 
nine are allowed by the PQ symmetry while the remaining nine, which do not appear in \eq{eq:11u}, 
are forbidden.  The PQ symmetry instead acts universally in the lepton sector so that 
all the Yukawa operators are permitted and there are no leptonic mixing effects. 
The structure of the operators in \eq{eq:11u} is enforced by the following set of charges 
\begin{align}
\X_{q_i} &= (2, 1, 0)\, ,\:
\X_{u_i} = (2 + s_\beta^2, s_\beta^2, - c_\beta^2)\, ,\:
\X_{d_i} = (2 + c_\beta^2, 1+c_\beta^2, c_\beta^2) \, , \nonumber \\
\X_{\ell_i}&= (1,1,1) s_\beta^2 \, ,
\end{align}
where $\tan\beta= \vev{H_2}/\vev{H_1} \equiv v_2/v_1$.
The numerical values of the non-vanishing Yukawa couplings that is 
required to reproduce the quark masses and CKM mixings (at the PQ-breaking scale) was computed in Ref.~\cite{Bjorkeroth:2018ipq}
\begin{align}
 y_u^{(1,1)} &= \frac{0.0009}{v_1}\, ,\: y_u^{(2,2)}=
   \frac{117.4}{v_2}\, ,\: y_u^{(3,2)}= 
   \frac{5.2\, +4.6 i}{v_1}\, ,\: y_u^{(3,3)}=
   \frac{0.4}{v_2}\, , \nonumber \\
     { y_d}^{(1,1)}&= \frac{0.0018}{ {v_2}}\, ,
     \: { y_d}^{(1,2)}=\frac{0.01}{ {v_1}}\, ,
     \: { y_d}^{(2,2)}=\frac{0.96}{ {v_2}}\, ,
     \: { y_d}^{(2,3)}=\frac{1.33}{ {v_1}}\, ,\: { y_d}^{(3,3)}= \frac{0.06}{ {v_2}} \, ,
   \end{align}
where (with a slight abuse of notation) $v_{1,2}$ in the denominators here stand for the 
dimensionless numbers $\vev{H_{1,2}}/$GeV. 
Diagonalization of the resulting mass matrices provides the numerical values 
of the LH and RH quark mixing matrices, so that the complete expressions of the 
axion couplings to the quarks can be readily obtained.  We have 
\begin{align}
&\frac{E}{N} =-\frac{10}{3} \, ,\quad 2N=1\,, \nonumber \\
&C_u =
s^2_\beta 
\, ,\quad 
C_c= -c^2_\beta-0.0036\, ,\quad
C_t= -c^2_\beta+0.0036\, , \nonumber \\ 
&C_d=c^2_\beta+ 0.1\, ,\quad 
C_s=c^2_\beta+0.55\, ,\quad 
C_b=c^2_\beta-0.66\, , \nonumber \\
&C_{e,\mu,\tau}=-s^2_\beta \, .  
\end{align}
The first  condition to realize nucleo-phobia is $C_u+C_d=1$~\cite{DiLuzio:2017ogq}
which is satisfied at the $10\%$ level. The second condition  
$C_u-C_d = 1/3$ is obtained for $\tan\beta=\sqrt{2}$ which is well within the 
perturbative Yukawa window 
$\tan\beta \in [0.117, 145]$.

\subsubsection*{$\mathcal{T}_2^{(d)}$}
\label{sec:M5}

The Yukawa sector of the $\mathcal{T}_2^{(d)}$ model contains a subset of nine quark operators, 
while for the leptons all Yukawa operators are allowed. They are: 
\begin{align}
& \bar q_1  u_1 H_1\, ,\ \:   \bar q_1  u_2 H_2\, ,\ \:\bar q_2 u_2 H_1\, ,
\ \:     \bar q_2  u_3 H_{2}\, ,\ \:     \bar q_3  u_3 H_{1}\, , \nonumber \\
  \label{eq:11d}
& \bar q_1  d_1 \tilde H_2\, , \ \:    \bar q_2  d_2 \tilde H_2\, ,  \ \:
  \bar q_3 d_3  \tilde H_{1}\, , \ \:    \bar q_3  d_2 \tilde H_{1}\, , \nonumber \\
  & \bar \ell_i e_j \tilde H_{2} \, .
\end{align}
The quark Yukawa structure in \eq{eq:11d} is enforced by the following set of charges:
\begin{align}
\X_{q_i} &= (2, 1, 0)\, ,\:
\X_{u_i} = (2 + s_\beta^2, 1 + s_\beta^2, 
    s_\beta^2)\, ,\: \X_{d_i} = (2 + c_\beta^2, c_\beta^2, 
   -s_\beta^2)\, ,\\
\X_{\ell_i} &=  -(1,1,1) c_\beta^2 \, .
\end{align}
The  values of the Yukawa couplings required to reproduce the quark  masses 
and CKM mixings
is~\cite{Bjorkeroth:2018ipq} 
\begin{align}
y_u^{(1,1)}&= \frac{0.0009}{v_1}\, ,\: y_u^{(1,2)}=
   \frac{1.1}{v_2}\, ,\:y_u^{(2,2)}=
   \frac{116.}{v_1}\, ,\:y_u^{(2,3)}=
   \frac{10.}{v_2}\, ,\:y_u^{(3,3)}= \frac{0.4}{v_1} \, , \nonumber \\
   y_d^{(1,1)}&= \frac{0.002}{v_2}\, ,\:y_d^{(2,2)}=
   \frac{1.7}{v_2}\, ,\:y_d^{(3,3)}=
   \frac{0.037}{v_1},\:y_d^{(3,2)}= \frac{0.06\, -0.03
   i}{v_1}\,.
\end{align}
where again $v_{1,2}$  stand for $\vev{H_{1,2}}/$GeV.
The axion couplings,  including   quark-mixing corrections  are:
\begin{align}
&\frac{E}{N} = \frac{20}{3}
\, ,\quad 2N=1\,,  \nonumber \\
&C_u =
s^2_\beta + 0.1
\, ,\quad 
C_c=-s^2_\beta -0.09\, ,\quad
C_t=-s^2_\beta - 0.01\, , \nonumber \\ 
&C_d=c^2_\beta\, ,\qquad \quad \ \ 
C_s=-s^2_\beta\, ,\qquad \qquad
C_b=-s^2_\beta\, , \nonumber \\
&C_{e,\mu,\tau}=c^2_\beta\, .
\end{align}
The first nucleo-phobic condition $C_u+C_d=1$ is satisfied at the $10\%$ level, while the second one $C_u-C_d = 1/3$ is obtained for $\tan\beta=\sqrt{2}$, well within the perturbative window
 which is well within the 
perturbative Yukawa window 
$\tan\beta\in [0.010,8.6]$.

\subsubsection{3HDM model}
\label{sec:3HDM}

A three Higgs doublets model (3HDM) which, besides nucleo-phobia, also allows to accomplish   
electro-phobic conditions (i.e.~approximate axion-electron decoupling)   was introduced in Ref.~\cite{Bjorkeroth:2019jtx}.
In this model the leptons couple to a third Higgs
doublet, $H_3$, with the same charges for all generations
\begin{align}
 \bar \ell_i e_j \tilde H_{3} \, .
\end{align}
In the quark sector the second generation replicates the PQ charges of the 
 first generation,  namely the structure is $2 + 1$ like in the $M_1$ model.
\begin{align}
&\bar q_1  u_1 H_1\, ,\ \:\bar q_3  u_3 H_{2}\, , \ \:\bar q_1  u_3 H_{1}\, , \ \:\bar q_3  u_1 H_{2}\, , \nonumber  \\
  \label{eq:3hdm}
&\bar q_1  d_1 \tilde H_2\, , \ \:\bar q_3  d_3 \tilde H_{1}\, ,\ \:\bar q_1  d_3 \tilde H_{2}\, ,\ \:\bar q_3  d_1 \tilde H_{1}\, .
\end{align}
For the scalar potential, the following couplings 
with the SM singlet scalar $\phi$ is assumed:
\begin{align}
H_3^\dagger H_1 \Phi^2 + H_3^\dagger H_2 \Phi^\dagger\,. 
\end{align}
These two operators fix  the PQ charges 
of $H_{1,2}$ in terms of the charge $\X_3$ of  $H_3$, as $\X_1=-2+\X_3$ and $\X_2 = 1+\X_3$.
The anomaly coefficients and the mixing independent part of  
the  couplings are
\begin{align}
&\frac{E}{N} = \frac{8}{3} 
,\quad 2N=1 \, , \nonumber \\
&c^0_{u,c} =\frac{2}{3} - \frac{\X_3}{3} \, , \quad 
c^0_t=-\frac{1}{3} - \frac{\X_3}{3} \, , \quad  \\
&c^0_{d,s}=\frac{1}{3} + \frac{\X_3}{3} \, , \quad 
c^0_b=-\frac{2}{3} + \frac{\X_3}{3} \, , \nonumber \\
&c^0_{e,\mu,\tau}=\frac{\X_3}{3} \, ,
\end{align}
and the condition of orthogonality between the PQ and the hypercharge scalar currents  
yields $\X_3 = 3(c_1^2 -1)c_2^2$  where $c_1 =\cos\beta_1$, $c_2 =\cos\beta_2$  
and the two vacuum angles are defined in terms of the doublet VEVs  $\vev{H_{1,2,3}}=v_{1,2,3}$ 
as $\tan\beta_1=v_2/v_1$ and $\tan\beta_2 = v_3/\sqrt{v_1^2+v_2^2}$. 
The values of the vacuum angles $\beta_{1,2}$  
are subject to the following perturbativity constraints~\cite{Bjorkeroth:2019jtx}:
\beq
\frac{y_t}{s_1 c_2} < \sqrt{\frac{16 \pi}{3}} \, , \qquad 
\frac{y_b}{c_1 c_2} < \sqrt{\frac{16 \pi}{3}} \, ,\qquad
\frac{y_\tau}{s_2} < \sqrt{4\sqrt{2}\pi} \, , 
\eeq
where $y_{t,b,\tau} = \sqrt{2} m_{t,b,\tau} / v$ are the heavy fermions Yukawa couplings 
expressed in terms of the quark masses and  $v \equiv (v_1^2+v_2^2+v_3^2 )^{1/2}  = 246$ GeV. \\

\begin{table}[t]
\footnotesize
\renewcommand{\arraystretch}{1.5}
\setlength\tabcolsep{3pt}
\begin{center}
\begin{tabular}{|c|c|c|c|c|c|c|c|c|}
\hline
Model & $E/N$ & $C_u$ & $C_d$ & $C_e$ & $C_c$ & $C_s$ & $C_t$ & $C_b$ \\
\hline
DFSZ1 & $8/3$ & $c^2_\beta/3$ & $s^2_\beta/3$ & $s^2_\beta/3$ & $c^2_\beta/3$ & $s^2_\beta/3$ & $c^2_\beta/3$ & $s^2_\beta/3$  
\\
\hline
DFSZ2 & $2/3$ & $c^2_\beta/3$ & $s^2_\beta/3$ & $-c^2_\beta/3$ & $c^2_\beta/3$ & $s^2_\beta/3$ & $c^2_\beta/3$ & $s^2_\beta/3$ 
\\
\hline
M1 & $2/3$ & $s^2_\beta$ & $c^2_\beta$ & $-s^2_\beta + \epsilon_L$ & $s^2_\beta$ & $c^2_\beta$ & $-c^2_\beta$ & $-s^2_\beta$ 
\\
\hline
M2 & $8/3$ & $s^2_\beta$ & $c^2_\beta$ & $-s^2_\beta + \epsilon_L$ & $s^2_\beta$ & $-s^2_\beta$ & $s^2_\beta$ & $-s^2_\beta$ 
\\
\hline
M3 & $-4/3$ & $c^2_\beta$ & $s^2_\beta$ & $s^2_\beta + \epsilon_L$ & $-s^2_\beta$ & $s^2_\beta$ & $-s^2_\beta$ & $s^2_\beta$ 
\\
\hline
M4 & $14/3$ & $s^2_\beta$ & $c^2_\beta$ & $-s^2_\beta + \epsilon_L$ & $s^2_\beta$ & $-s^2_\beta$ & $s^2_\beta$ & $-s^2_\beta$ 
\\
\hline
$\mathcal{T}_2^{(u)}$ & $-10/3$ & $s^2_\beta$ & $c^2_\beta +0.1$ & $-s^2_\beta$ & $-c^2_\beta - 0.0036$ & $c^2_\beta +0.55$ & $-c^2_\beta + 0.0036$ & $c^2_\beta -0.66$ 
\\
\hline
$\mathcal{T}_2^{(d)}$ & $20/3$ & $s^2_\beta +0.1$ & $c^2_\beta$ &  $c^2_\beta$ & $s^2_\beta -0.09$ & $-s^2_\beta$ & $s^2_\beta -0.01$ & $-s^2_\beta$ 
\\
\hline
3HDM & $8/3$ 
& $\frac{2}{3} - \frac{\X_3}{3}$ 
& $\frac{1}{3} + \frac{\X_3}{3}$
& $\frac{\X_3}{3}$ 
& $\frac{2}{3} - \frac{\X_3}{3}$  
& $\frac{1}{3} + \frac{\X_3}{3}$ 
& $-\frac{1}{3} - \frac{\X_3}{3}$ 
& $-\frac{2}{3} + \frac{\X_3}{3}$
\\
\hline
\end{tabular}
\caption{Summary of the relevant axion-fermion couplings and $E/N$ values for the various models.
For the meaning of the leptonic mixing parameter $\epsilon_L$ see text. 
\label{tab:axionmodels}
}
\end{center}
\end{table}

\subsection{Summary of axion models and discussion}
\label{sec:pert}

The values of $E/N$ and of the axion-fermion couplings relevant
for the astrophysical analysis 
for all the models reviewed in this Section are collected in~\Table{tab:axionmodels}.

In general, in models with non universal PQ charge assignments,   axion interactions 
with matter fields will feature a certain amount of flavour violation. 
Searches for kaon decays of the type $K\to \pi a$ provide the strongest constraints 
on flavour-violating axion couplings (see for example 
Table~2 in Ref.~\cite{Bjorkeroth:2018dzu}). 
For model M3 the charge assignments for the quark doublets and for the RH down type quarks  respect universality, so this model automatically evades the corresponding limits.
Model M1 features universality of the couplings for the RH quarks of the three generations, 
but for the LH quark doublets only the first two families have equal charges. 
In this case, if the off-diagonal entries $V^d_{13},V^d_{23}$ in the LH mixing matrix  are 
not particularly suppressed, the predicted $K\to \pi a$  decay rate could easily 
conflict with  the experimental limit. If instead the mixings are at 
most CKM-like ($V^d_{13},V^d_{23}\ll 1$) model M1 would also evade the $K\to \pi a$ constraints.
Because of the unitarity constraints on the mixing matrices, this also ensures that the diagonal couplings
of the axion to the first generation quarks would be  negligibly affected so that, to a good approximation, 
the conditions for axion nucleo-phobia remain preserved.
Models M2 and M4 feature 
universality for the LH quark doublets and RH up-type quarks, however 
in the RH down-sector the charges of the first and second generation differ. In this case 
even for  CKM-like RH mixings the decay rate would exceed the experimental limits by 
several orders of magnitude~\cite{DiLuzio:2017ogq}. Nevertheless, as it was shown 
in Ref.~\cite{Bjorkeroth:2018ipq}, it is possible to assume some specific 
mass-matrix textures  in such a way that, for example, $s_R$ 
does not mix with the other RH down-type quarks, in which case 
$K\to \pi a$ (and also $B\to K a)$
limits are easily evaded. 
For example,  the  1+1+1 models $\mathcal{T}^{u,d}_2$ 
incorporate precisely this feature (cf.~Ref.~\cite{Bjorkeroth:2018ipq}).
Although in these models the PQ charges 
of all the three generations differ, it is the PQ symmetry itself 
that enforces suitable matrix textures such that dangerous off-diagonal mixings are absent and  
flavour-diagonal axion couplings to the quarks of the first generation  are  negligibly affected. 
Hence flavor violation limits can be evaded 
while preserving the nucleo-phobic property of the axion.

A second issue regards the range in which the vacuum angles that define 
the couplings of the physical axion are allowed to vary, and that should respect 
the perturbativity constraints on the relevant Yukawa couplings. 
In order to estimate the perturbativity domain 
we have employed the tool of perturbative 
unitarity on the $2 \to 2$ scatterings of the Yukawa theory in the 
presence of two or more Higgs doublets. 
In particular, taking into account gauge group factors (see e.g.~\cite{DiLuzio:2016sur,DiLuzio:2017chi,Allwicher:2021jkr})
one gets \cite{Bjorkeroth:2019jtx}: 
$y_q < \sqrt{16 \pi / 3}$ for the Yukawa couplings of the quarks and 
$y_\ell < \sqrt{4\sqrt{2} \pi}$ for the charged leptons, where  $y_{q,\ell}$ are 
the Yukawa couplings defined in the multi-Higgs doublet  
theory that have to be 
matched with their SM counterparts in order to extract a bound on the 
electroweak vacuum angles. This matching presents some model dependency
related to the scale at which the extra Higgs doublets are integrated 
out, 
which can vary between the PQ and the electroweak scale. 
For DFSZ1-2, M1-2-3-4 and 
3HDM models we perform the matching at the electroweak scale (which allows to neglect top running effects in the  
low-energy axion couplings~\cite{Choi:2017gpf,Choi:2020rgn,Chala:2020wvs,Bauer:2020jbp,Choi:2021kuy}), 
while for $\mathcal{T}_2^{(u,d)}$ models which feature 
textures in the quark mass matrices 
that are enforced by the PQ symmetry, 
we perform the matching at the PQ scale,  
and we neglect for simplicity top-related running effects.

\section{Axion models in the light of cooling anomalies}
\label{sec:fits}

This Section is devoted to the interpretation of the cooling anomalies discussed in Sec.~\ref{sec:astroobs} in a model-dependent way, putting under scrutiny the axion models presented in Sec.~\ref{sec:models}. 
Specifically, the results of the fits to the astrophysical data will be presented in Sec.~\ref{sec:cooling fits}, where the ability to reproduce data will be scrutinized and the models capable to provide a better accordance with observations will be singled out. Then, in Sec.~\ref{sec:haloscopes}, we will analyze the experimental potential to probe these models in the region relevant for stellar evolution. 

\subsection{Axion fits to cooling anomalies}
\label{sec:cooling fits}

In order to carry out our Bayesian analyses, we implemented all the models under investigation in the public \texttt{HEPfit} package~\cite{deBlas:2019okz}, which performs a Markov Chain Monte Carlo (MCMC) analysis by means of the Bayesian Analysis Toolkit (BAT)~\cite{Caldwell:2008fw}. This framework implements a Metropolis-Hastings algorithm, with the MCMC runs involving 20 chains with a total of $10^8$ events per run, collected after all chains have achieved convergence in an adequate number of pre-run iterations.

Before performing the runs, we assigned a theoretical prior to each of the model parameters entering the fits. In particular, $\tan \beta$ is scanned logarithmically, in a similar fashion to what done in Ref.~\cite{Hoof:2018ieb}, in the broad range $0.01$-$10^3$.
For the axion mass we also performed a logarithmic scan, choosing the range  $0.1$-$10^3$~meV which corresponds to the mass region of phenomenological interest. The mixing correction to the electron coupling $\epsilon_L$, which is not determined by the theory and in the  2+1 models represents an additional free parameter, is assumed to be distributed flatly in the physical range~$[-1,1]$. Similarly, the $\X_3$ parameter of 3HDM is not constrained and is flatly distributed in the range~$[0,1]$.

The Bayesian model comparison between the different models 
is performed evaluating for each of them an approximation of the Bayes factor, namely the \textit{Information Criterion} (IC)~\cite{IC}.\footnote{This quantity provides an approximation for the predictive accuracy of a model, and is defined from the mean and the variance of the posterior probability distribution function (\emph{p.d.f.}) of the log-likelihood $\log \mathcal{L}$, $\text{IC} \equiv -2 \overline{\log \mathcal{L}} \, + \, 4 \sigma^{2}_{\log \mathcal{L}}$.}
When comparing two models, the preferred one is the model displaying the smallest IC value~\cite{2013arXiv1307.5928G}. Hence, after computing the IC value of the SM, we  compute for each model $\mathcal{M}$ the score factor $\Delta \text{IC} \equiv \text{IC}_{\rm SM} - \text{IC}_\mathcal{M}$, which is a quantity expressed in units of standard deviations. A larger value for this score thus signals a greater improvement in  reproducing the data,  compared to the SM case. The preferred models are thus the ones displaying the higher values of $\Delta \text{IC}$.

The results of the global fits are reported in Table~\ref{tab:fit_results}. Remembering that the constrains stemming from data relative to SN 1987A and to various NS are less sound than the ones extracted from WD, RGB and HB, and considering that, as we will see, they do affect in a strong manner the outcome of the fit, we report the results obtained for each model in three different analyses: in the first, less inclusive one, only data from WD, RGB and HB are included; in the second, more inclusive one, we fitted also the data pertaining to SN 1987A; and in the last, 
the most inclusive one, both data sets stemming 
from SN 1987A and  NS are  added to the likelihood. For each model and for each set of observables included in the fit we list the values of its parameters at the global mode, corresponding to the point where the multi-dimensional posterior \emph{p.d.f.}
is maximal. Only for the less inclusive fit for DFSZ1, for which the global mode of a parameter would fall  outside the range of perturbative unitary discussed in Sec.~\ref{sec:models}, we imposed 
as an additional prior also a unitarity constraint. This procedure is however employed solely for the sake of reporting in Table~\ref{tab:fit_results} a value at the global mode within the unitarity limits, while all the remaining information pertaining to the model are inferred without imposing it. Finally, for each fit we give in the last column the value of the score factor $\Delta \text{IC}$. 

\begin{table}[!th!]
\renewcommand{\arraystretch}{1.5}
\begin{center}
\footnotesize
\begin{tabular}{|c|c||c|c|c|c||c|}
\hline
Model & Obs. included in the fit & $m_a$ [meV] & $\tan \beta$ & $\epsilon_L$ & $\X_3$ & $\Delta$IC \\
\hline \multirow{3}{*}{DFSZ1} &
WD, RGB, HB &
79 & 0.250 & - & - & 2.0 \\ 
& WD, RGB, HB, SN &
12.1 & 0.802 & - & - & 1.6 \\ 
& WD, RGB, HB, SN, NS &
11.1 & 0.807 & - & - & 0.9 \\ 
\hline \multirow{3}{*}{DFSZ2} &
WD, RGB, HB &
68 & 4.04 & - & - & 0.9 \\ 
& WD, RGB, HB, SN &
8.4 & 0.954 & - & - & 0.7 \\ 
& WD, RGB, HB, SN, NS &
6.6 & 0.681 & - & - & 0.5 \\ 
\hline \multirow{3}{*}{M1} &
WD, RGB, HB &
97 & 1.60 & \phantom{-}0.81 & - & 4.7 \\ 
& WD, RGB, HB, SN &
77 & 1.50 & \phantom{-}0.76 & - & 4.3 \\ 
& WD, RGB, HB, SN, NS &
53 & 1.37 & \phantom{-}0.73 & - & 2.6 \\ 
\hline \multirow{3}{*}{M2} &
WD, RGB, HB &
123 & 1.81 & \phantom{-}0.75 & - & 4.7 \\ 
& WD, RGB, HB, SN &
88 & 1.50 & \phantom{-}0.71 & - & 4.2 \\ 
& WD, RGB, HB, SN, NS &
55 & 1.40 & \phantom{-}0.70 & - & 2.6 \\ 
\hline \multirow{3}{*}{M3} &
WD, RGB, HB &
63 & 0.585 & -0.27 & - & 4.9 \\ 
& WD, RGB, HB, SN &
56 & 0.569 & -0.27 & - & 4.6 \\ 
& WD, RGB, HB, SN, NS &
51 & 0.584 & -0.27 & - & 2.9 \\ 
\hline \multirow{3}{*}{M4} &
WD, RGB, HB &
79 & 1.70 & \phantom{-}0.81 & - & 4.8 \\ 
& WD, RGB, HB, SN &
63 & 1.49 & \phantom{-}0.76 & - & 4.5 \\ 
& WD, RGB, HB, SN, NS &
50 & 1.38 & \phantom{-}0.73 & - & 2.7 \\ 
\hline \multirow{3}{*}{$\mathcal{T}_2^{(u)}$} &
WD, RGB, HB &
63 & 0.125 & - & - & 3.1 \\ 
& WD, RGB, HB, SN &
13 & 0.398 & - & - & 2.9 \\ 
& WD, RGB, HB, SN, NS &
5.0 & 0.707 & - & - & 2.2 \\ 
\hline \multirow{3}{*}{$\mathcal{T}_2^{(d)}$} &
WD, RGB, HB &
67 & 7.08 & - & - & 2.0 \\ 
& WD, RGB, HB, SN &
25 & 3.54 & - & - & 1.9 \\ 
& WD, RGB, HB, SN, NS &
10 & 2.51  & - & - & 1.2 \\ 
\hline \multirow{3}{*}{3HDM} &
WD, RGB, HB &
103 & - & - & 0.03 & 3.7 \\ 
& WD, RGB, HB, SN &
97 & - & - & 0.04 & 3.3 \\ 
& WD, RGB, HB, SN, NS &
94 & - & - & 0.04 & 1.7 \\ 
\hline
\end{tabular}
\end{center}
\caption{Values of the parameters at the global mode found by the fits. For each model we give the results of three different fits performed by including in the likelihood three different sets of observables. In the last column, we list as a goodness-of-fit measurement 
relative to the SM case, the values of $\Delta$IC in units of standard deviations. 
The higher the value of $\Delta $IC the  better the agreement with the data.}
\label{tab:fit_results}
\end{table}

Starting from the canonical DFSZ models, we observe that the results are qualitatively similar to the ones found in the previous global analysis of Ref.~\cite{Giannotti:2017hny}. Indeed, when considering only data from WD, RGB and HB, we obtain a better agreement with data compared to the SM, with the global mode for the axion mass found at $m_a \sim$ 70-80 meV. On the other hand, including also data  from SN 1987A and from the NS has the well understood effect of strongly reducing the allowed mass range for the axion, with the global mode now sitting at $m_a \sim$ 10 meV. Given the fact that these additional measurements only put an upper bound on the axion couplings to nucleons, without hinting to new physics, the SM without axions is already in compliance with them.
This is the reason why the overall preference for  DFSZ models (or for any kind of axion model) over the SM is reduced when considering all experimental information at hand. Nevertheless, due to the presence of cooling anomalies, the preference 
for   DFSZ models over the SM persists even in the most inclusive case, with DFSZ1 
performing slightly better than DFSZ2.

Moving now to the non-universal cases, we start our discussion from the 2+1 models. 
Among all the models that we have  considered these are the ones that display the better agreement with data. Thanks  to their nucleo-phobic character, this feature is maintained even when observations stemming from SN 1987A and the NS are included in the fit. As a consequence,
the global modes of the axion mass, which in the less inclusive fit  for the M2 model 
was found to be as high as $m_a \sim$ 123 meV, 
do not drop below $m_a \sim$ 50 meV even in the most inclusive cases. An interesting feature that can  be inferred from~\Table{tab:axionmodels} is that the models M1, M2 and M4 share many similarities: the global modes of the lepton mixing parameter are always found at $\epsilon_L \sim 0.75$, while the inferred value for the M3 model is $\epsilon_L \sim 0.27$; the global mode of $\tan\beta$ follows an analogous pattern, sitting at $\tan\beta \sim 1.5$ for the former models and 
at $\tan\beta \sim 0.5$ for the latter. 
Moreover, it is also evident 
that when adding SN and NS observables the global mode approaches the nucleo-phobic point 
$\tan\beta = \sqrt{2}$ (for M2, M3 and M4) and $\tan\beta = 1/\sqrt{2}$ 
for M3.

The fits for the 1+1+1 models $\mathcal{T}_2^{(u,d)}$ share some similarities with the universal DFSZ models, with the global mode for the axion mass ranging from $m_a \sim$ 65 meV in the less inclusive fits to $m_a \sim$ 10 meV in the more inclusive ones. However, given the nucleo-phobic nature of these models their overall agreement with data is improved compared to the universal models, hence suggesting a preference of $\mathcal{T}_2^{(u,d)}$ over DFSZ1 and DFSZ2, even if less strong than for the 2+1 models.

The last model analysed in our study is  the 3HDM. In the corresponding fits the global mode for the axion mass is particularly stable, being found at $m_a \sim$ 103 meV in the less inclusive fit and at $m_a \sim$ 94 meV in the most inclusive one. This behaviour is due to the astro-phobic nature of the model: that is the nucleo-phobic conditions can be satisfied while simultaneously suppressing the axion-electron couplings (the global mode for the electron charge $\X_3$ is always found at $\X_3 \sim$ 0.04 well within  the electro-phobic regime). For this reason the results of the fit are not strongly affected by increasing the number of observables included 
in the likelihood. Similarly to $\mathcal{T}_2^{(u,d)}$, 
the 3HDM model is preferred by data over DFSZ1 and DFSZ2, but again not as strongly as the 2+1 models.

\begin{figure}[!t]
\centering
\includegraphics[width=0.37\textwidth]{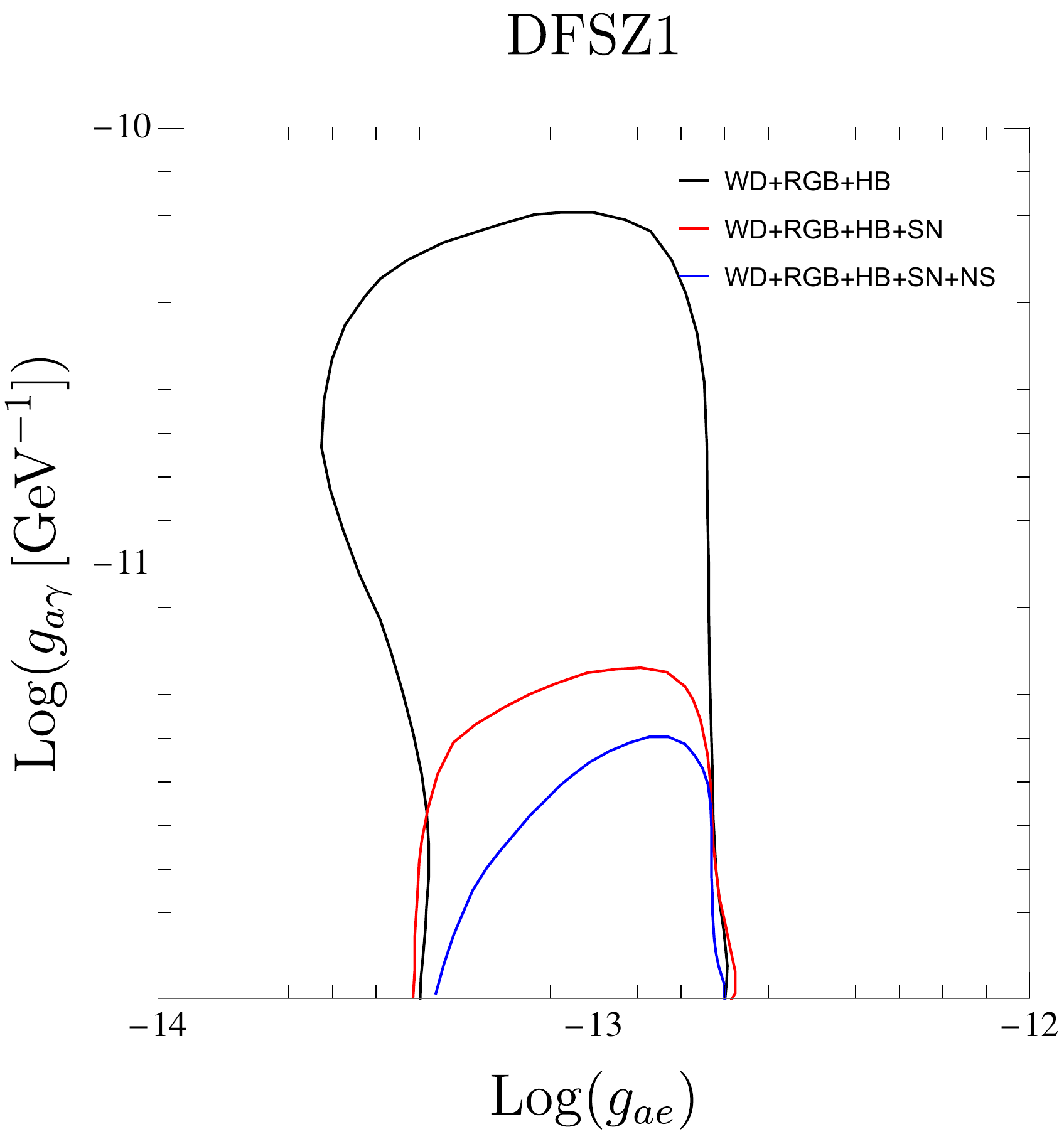}
\includegraphics[width=0.37\textwidth]{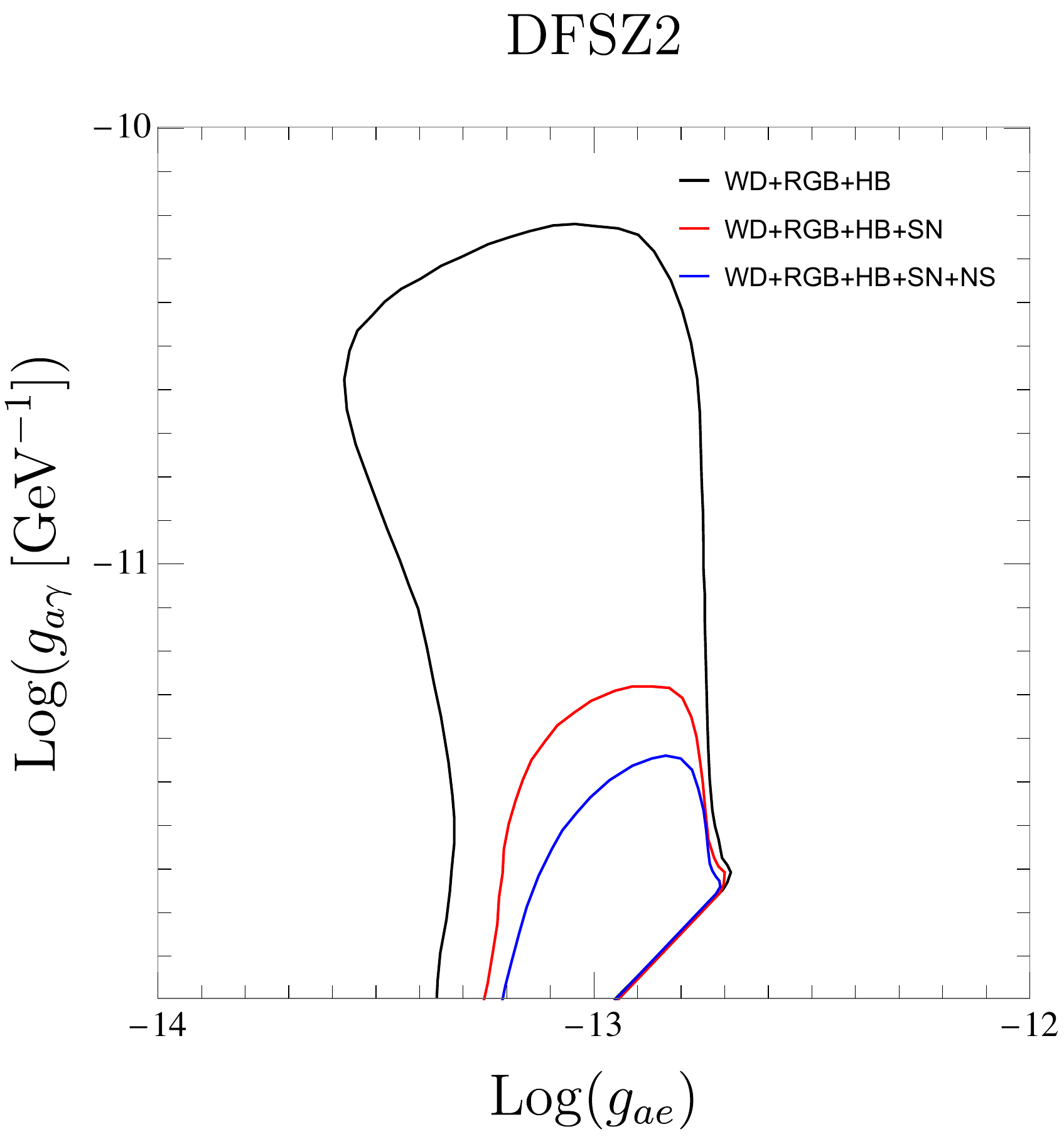}
\includegraphics[width=0.37\textwidth]{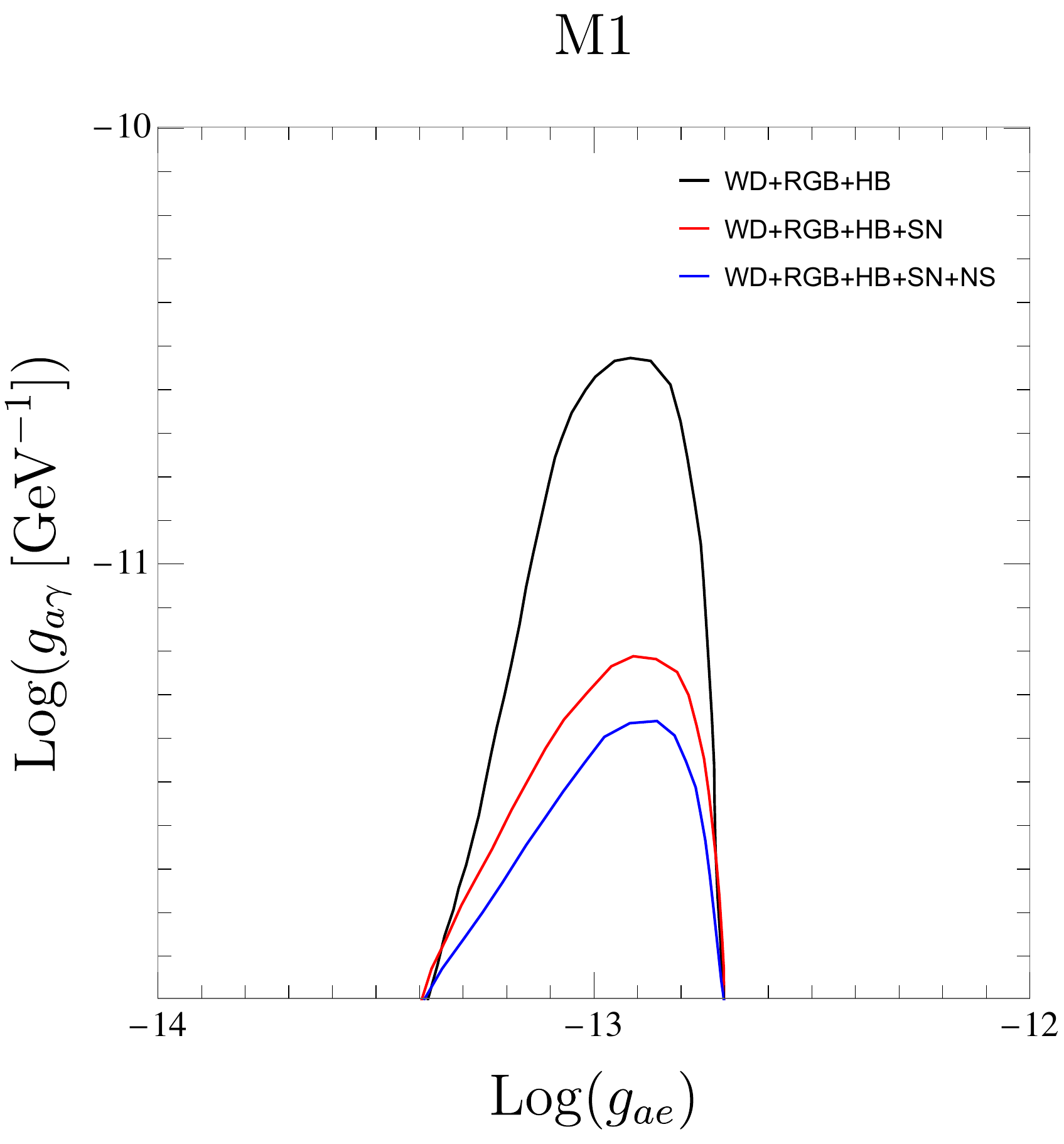}
\includegraphics[width=0.37\textwidth]{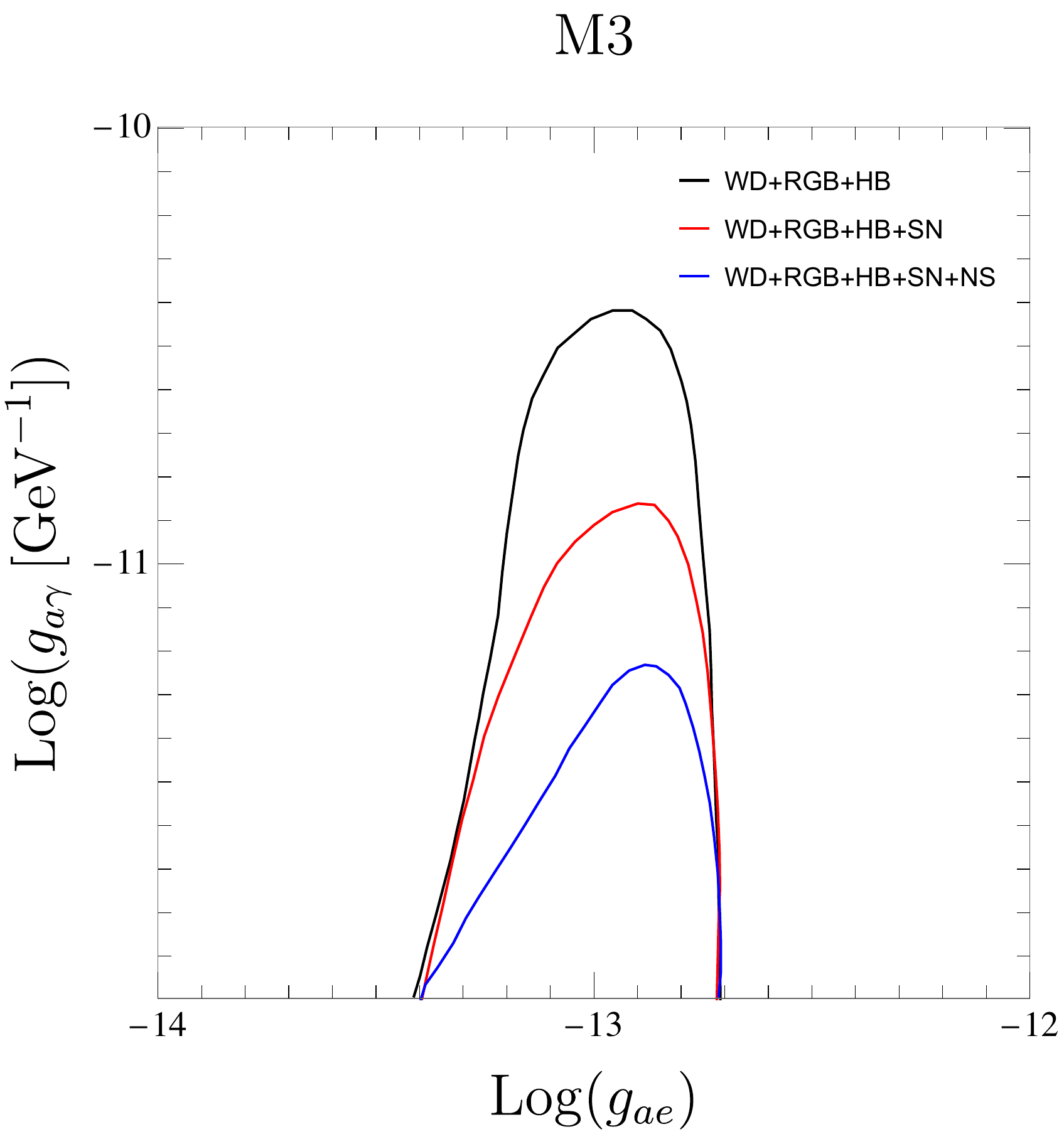}
\includegraphics[width=0.37\textwidth]{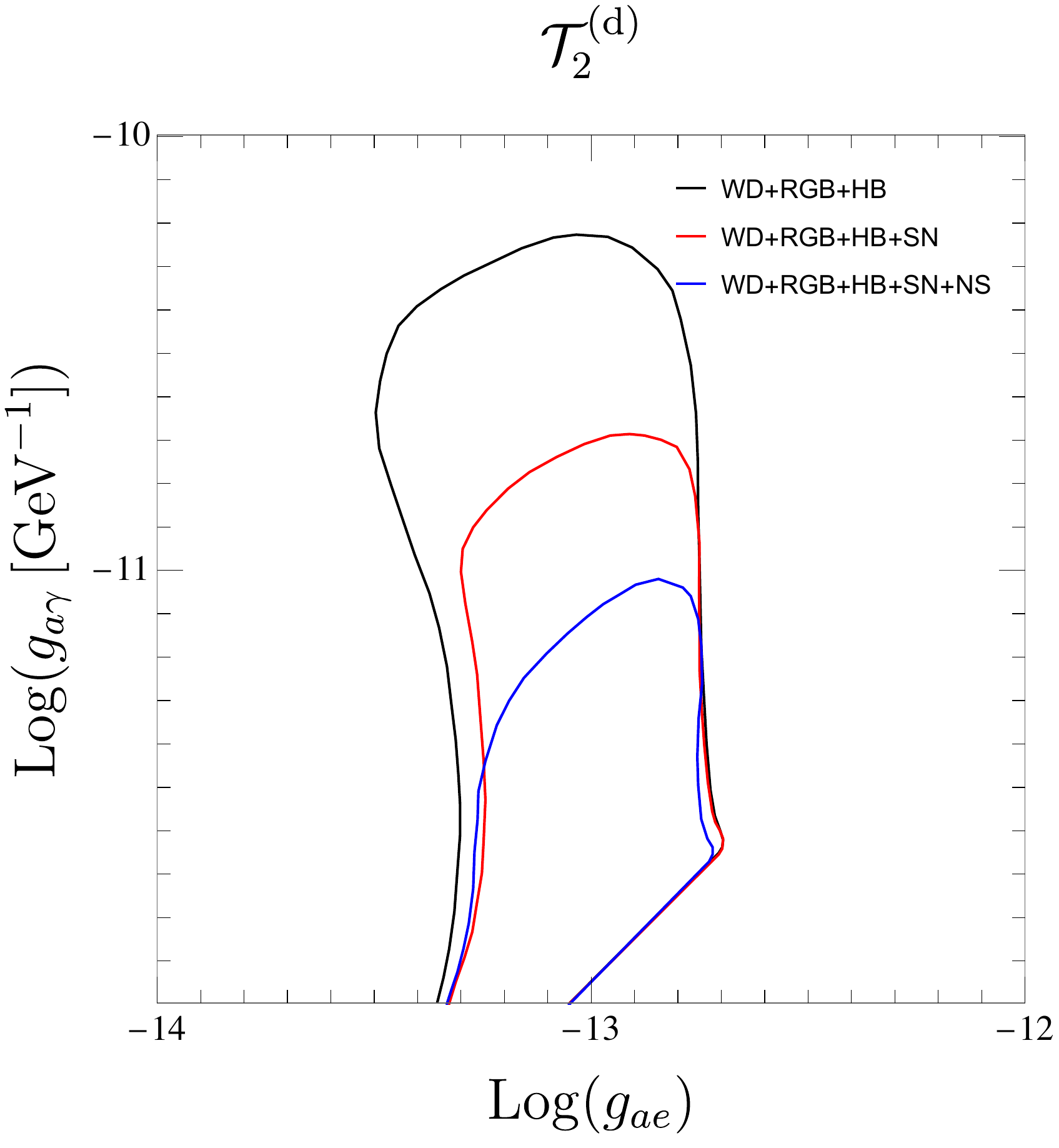}
\includegraphics[width=0.37\textwidth]{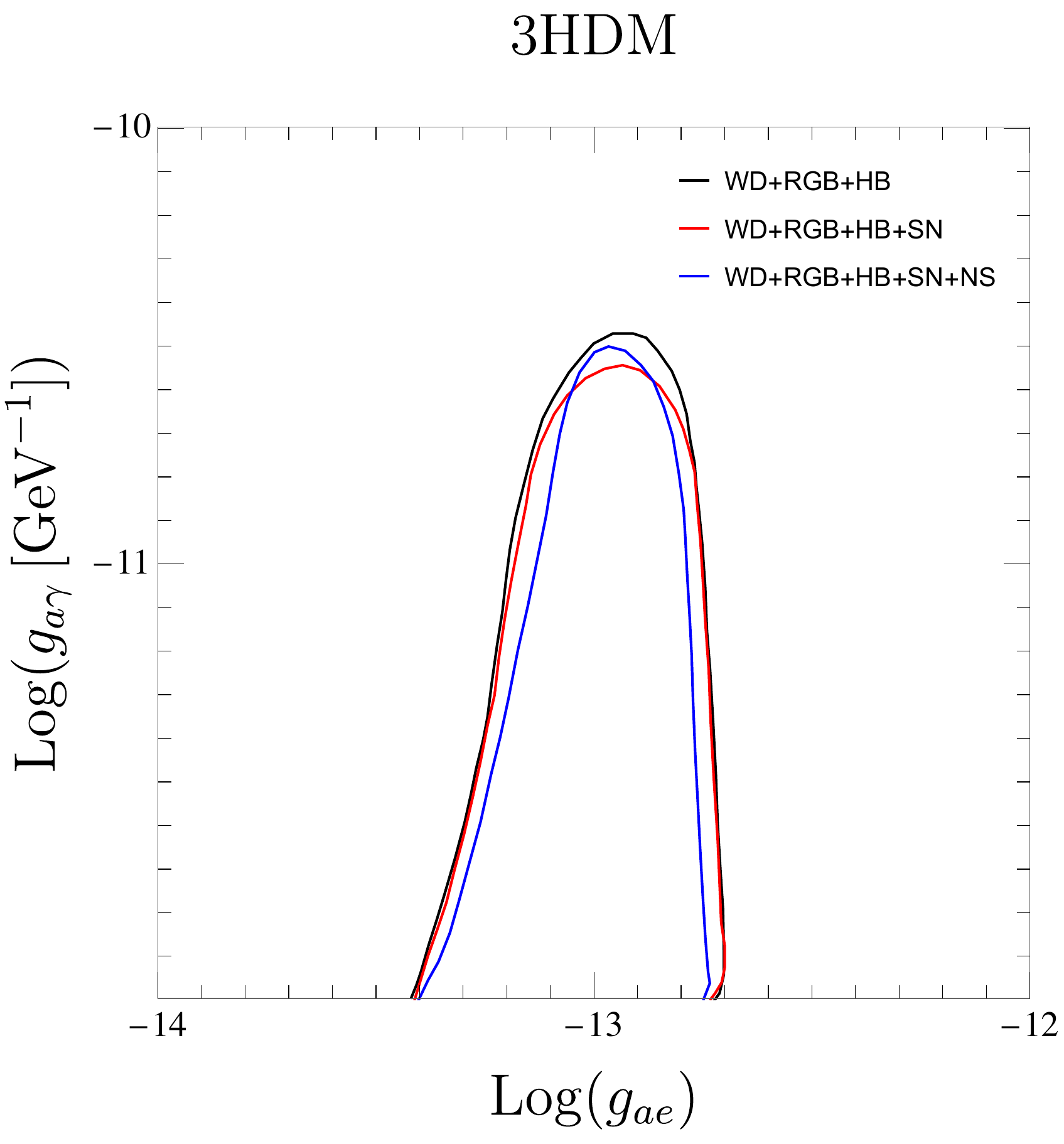}
\caption{Contours of the $2\sigma$ regions from global fits to stellar cooling in the $g_{ae}$ vs. $g_{a\gamma}$ plane for different models. We do not show the results for models M2 and M4 that are qualitatively similar to the ones of the model M1, and the ones relative to $\mathcal{T}_2^{(u)}$, which strongly resemble the ones of model $\mathcal{T}_2^{(d)}$. Black lines correspond to the less inclusive global fit with only WD, RGB and HB data; red lines correspond to the more inclusive fit with data from SN 1987A added; blue lines are relative to the most inclusive case in which also data from NS are taken into account.}
\label{fig:gaegag_md}
\end{figure}

Complementary to Table~\ref{tab:fit_results}, we also give in Fig.~\ref{fig:gaegag_md} the $2\sigma$ bounds extracted from the fits in the $g_{ae}$ vs. $g_{a\gamma}$ plane. For each model we show those regions for 3 different cases, according to the different set of observables included in the analysis.
We recall that, contrarily to the model-independent case studied in Sec.~\ref{sec:modindepfit} where 
$g_{ae}$ and $g_{a\gamma}$ are considered as independent parameters, here these 
two couplings depend on the same model parameters and  hence are strongly correlated. This is the reason why the panels in Fig.~\ref{fig:gaegag_md} are somewhat different compared to the plot shown in Fig.~\ref{fig:gaegag}, particularly concerning $g_{a\gamma}$ and when observables constraining the couplings to the nucleons are included in the fit. Indeed, as we have discussed above, the data relative to SN 1987A and the NS generally lower the upper bound on $m_a$, which in turn decreases the upper bound for the coupling $g_{a\gamma}$. This effect is more evident for the universal DFSZ models, but less accentuated for the non-universal ones. It is also worth recalling that, while 
in non universal 2+1 and 1+1+1 models the nucleo-phobic conditions are obtained only for a specific value of the free parameter $\tan\beta$, they are always realised in the 3HDM model. This is the reason why the results of the latter model remain stable against the inclusion in the fit of the  SN 1987A and of the  NS data. 
As a closing note, it is intriguing to observe that the values of the axion mass 
which better fit the astrophysical data
reported in Table~\ref{tab:fit_results}, 
fall in the 
range 0.2-100 meV that is 
suggested, in post-inflationary scenarios, by cosmological data~\cite{Hoof:2021jft}.

\subsection{Discovery potential of meV-scale axion experiments}
\label{sec:haloscopes}

In this Section we discuss the perspectives  to access experimentally the 
parameter regions hinted by the cooling anomalies.

Fig.~\ref{fig:gaegag_md} shows that all models require a finite coupling to electrons to explain the observed stellar behavior.
It is tempting, therefore, to look at experiments sensitive to the axion-electron coupling 
to test the preferred  regions for this parameter.
The most sensitive experiments of this kind are the large underground detectors XENON1T~\citep{XENON:2020rca}, LUX~\citep{Akerib:2017uem} and 
PandaX-II~\citep{PandaX:2017ock}.
All of them can search for solar axions converted in the detector through the axio-electric effect~\cite{Arisaka:2012pb}. 
This strategy, however, has so far allowed to probe only relatively large axion-electron couplings, 
in a region in tension with RGB and WD observation (see, e.g., Ref.~\cite{DiLuzio:2020jjp}).
Even the next generation of underground detectors, such as Darwin~\cite{DARWIN:2016hyl}, will not have a sufficient sensitivity to 
probe the couplings favored by stellar evolution (see, e.g., Ref.~\cite{Irastorza:2018dyq,Baudis:2021ipf}).
As we shall see, probing the axion-photon coupling offers a more efficient way to dig into the parameter region preferred by stars. 
Indeed, Tab.~\ref{tab:fit_results} shows that  
all our representative models give a maximal agreement with the stellar observations for 
axion masses in a range from a few meV to $\sim$ 100 meV,
a mass range accessible to 
the next generation of axion helioscopes. 
As the name indicates,  helioscopes~\cite{Sikivie:1983ip} search for axions produced in the Sun.   
In the Sun core the main axion production channels are the Primakoff and the ABC (atomic transitions, bremsstrahlung and Compton) processes.
The solar axion flux on earth is (see, e.g., Ref.~\cite{DiLuzio:2020jjp}),
\begin{align}\label{eq:solar_spectrum_integrated}
\frac{dN_a}{dt}
=1.1\times 10^{39} 
\left[ 
\left( 
\frac{g_{a\gamma}}{10^{-10}{\rm GeV}^{-1}} 
\right)^2  
+ 0.7\,
\left( 
\frac{g_{ae}}{10^{-12}} 
\right)^{2} 
\right] \,{\rm s}^{-1}  \,.
\end{align}
We notice that this flux gets a similar contribution from Primakoff and ABC  for values 
of the couplings of phenomenological interest.
The exact weight of the two contributions depends on the specific axion model.
There are additional contributions to the solar flux induced by the axion-nucleon coupling.
These are, however, normally peaked at energies too large for standard helioscopes and will be ignored here.\footnote{A notable exception is the decay of $^{57}$Fe~\cite{Avignone:2017ylv}, with the emission of a narrow 14.4 keV axion line.  
This flux was already searched a decade ago by the CAST helioscope~\cite{CAST:2009jdc}.
However, 
the axion flux from the decay of $^{57}$Fe is normally sub-leading  with respect  to the other contributions.
Furthermore, 
it is not yet clear if BabyIAXO or IAXO will
be optimized for a detection at such high energies, where the X-ray optics might be less efficient. 
Because of these reasons, the $^{57}$Fe contribution to the axion flux will not be considered further in the present discussion. }
 
Helioscopes exploit a strong laboratory magnetic field to convert  solar axions into X-ray photons. 
The conversion probability is 
\begin{align}
\label{eq:Helioscope_agamma_probability}
P_{a\to \gamma}=\left( \dfrac{g_{a\gamma}\, B\, L}{2}\right)^2 \dfrac{{\rm sin}^2(qL/2)}{(qL/2)^2}\,,
\end{align}
where $ q=q_\gamma-q_a $ is the momentum transfer provided by the magnetic field and $L$ is the length of the magnet.
In vacuum,  $ q\simeq m_a^2/2\omega $.
Coherence is ensured when $qL\ll 1$.
Therefore, the helioscope sensitivity is practically mass-independent up to a certain mass threshold, $m_{\rm th}$, above which it drops rapidly. 
The mass threshold depends on the specific helioscope. 
From the above expression, we find 
$m_{\rm th}\approx 10\, {\rm meV}\,L_{10}^{-1/2}$, 
where $L_{10}$ is the magnet length in units of 10 m and we are using 3 keV as a reference solar axion energy. 
Whenever the coherence condition is realised, the conversion probability scales as $(g_{a\gamma}BL)^2$, rapidly increasing with the magnetic field and with the size of the magnet. 
Above $m_{\rm th}$, the sensitivity is reduced proportionally to $m_a^{-2}$.
The sensitivity may be regained using a buffer gas in the magnet beam pipes~\cite{vanBibber:1988ge}.
In this case the momentum transfer is $ q\simeq (m_a^2-m_\gamma^2)/2\omega $. 
Tuning the effective photon mass, $m_\gamma$, to the axion mass allows to effectively regain coherence.

Below, we present a detailed study of the potential of the next generation of axion helioscopes to 
probe, in the regions of astrophysical interest, the various models that we have been discussing.  In particular, we consider the proposed International Axion Observatory (IAXO)~\cite{Irastorza:2011gs,Armengaud:2014gea, Giannotti:2016drd,IAXO:2019mpb}, 
and its scaled versions BabyIAXO and IAXO+. 
Of particular interest is BabyIAXO, a scaled down (and significantly  less expensive) version of IAXO, 
which will likely start operations in the mid of the current decade
at the Deutsches Elektronen-Synchrotron (DESY)~\cite{BabyIAXO:2020mzw, Abeln:2020eft}.
The beginning of operations for IAXO are presently not known.
However, even this larger helioscope does not present particular technical challenges, 
besides better components~\cite{IAXO:2019mpb} and may become operational in a not so distant future.
\begin{figure}[!th]
\centering
\includegraphics[width=0.37\textwidth]{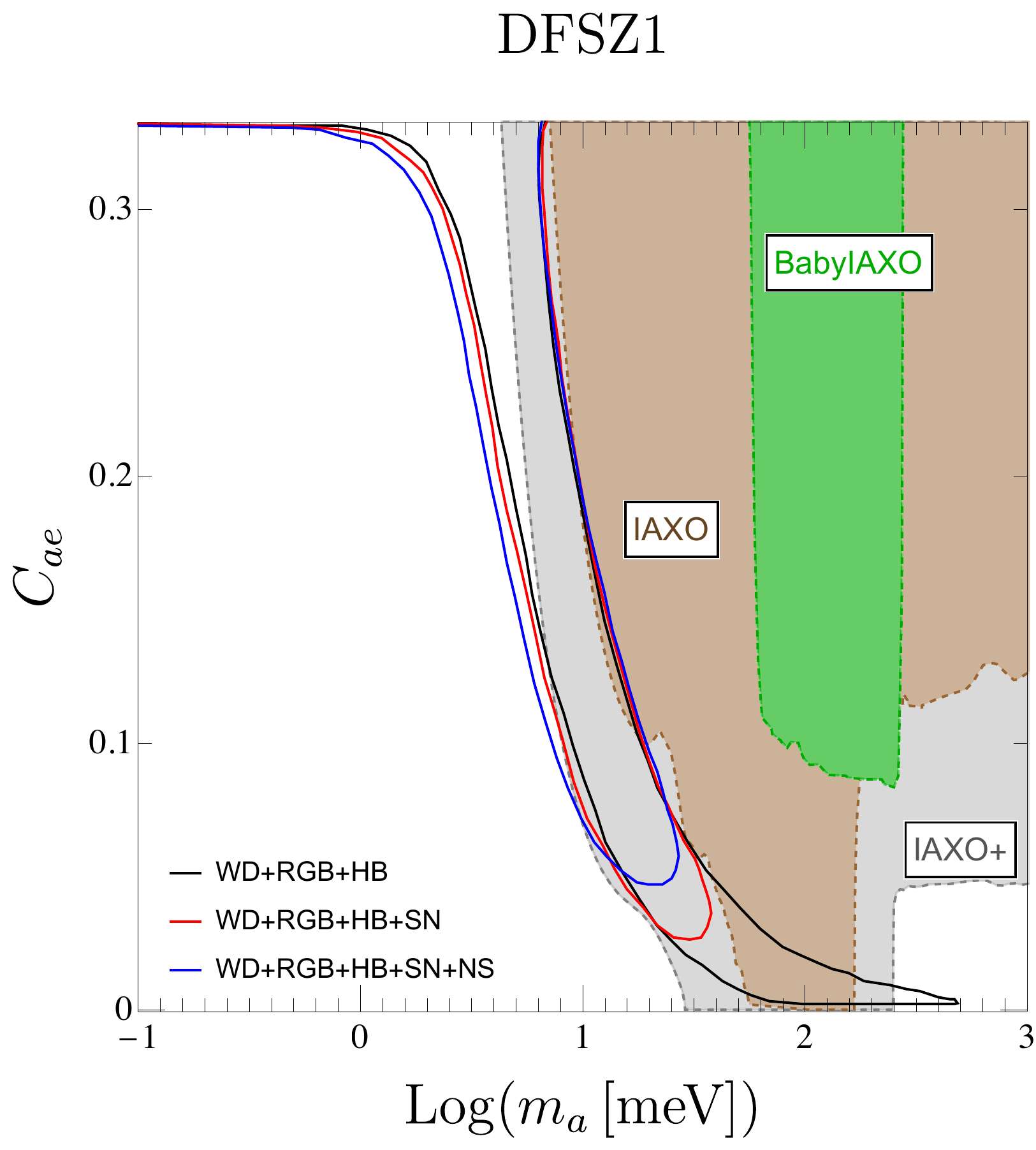}
\includegraphics[width=0.37\textwidth]{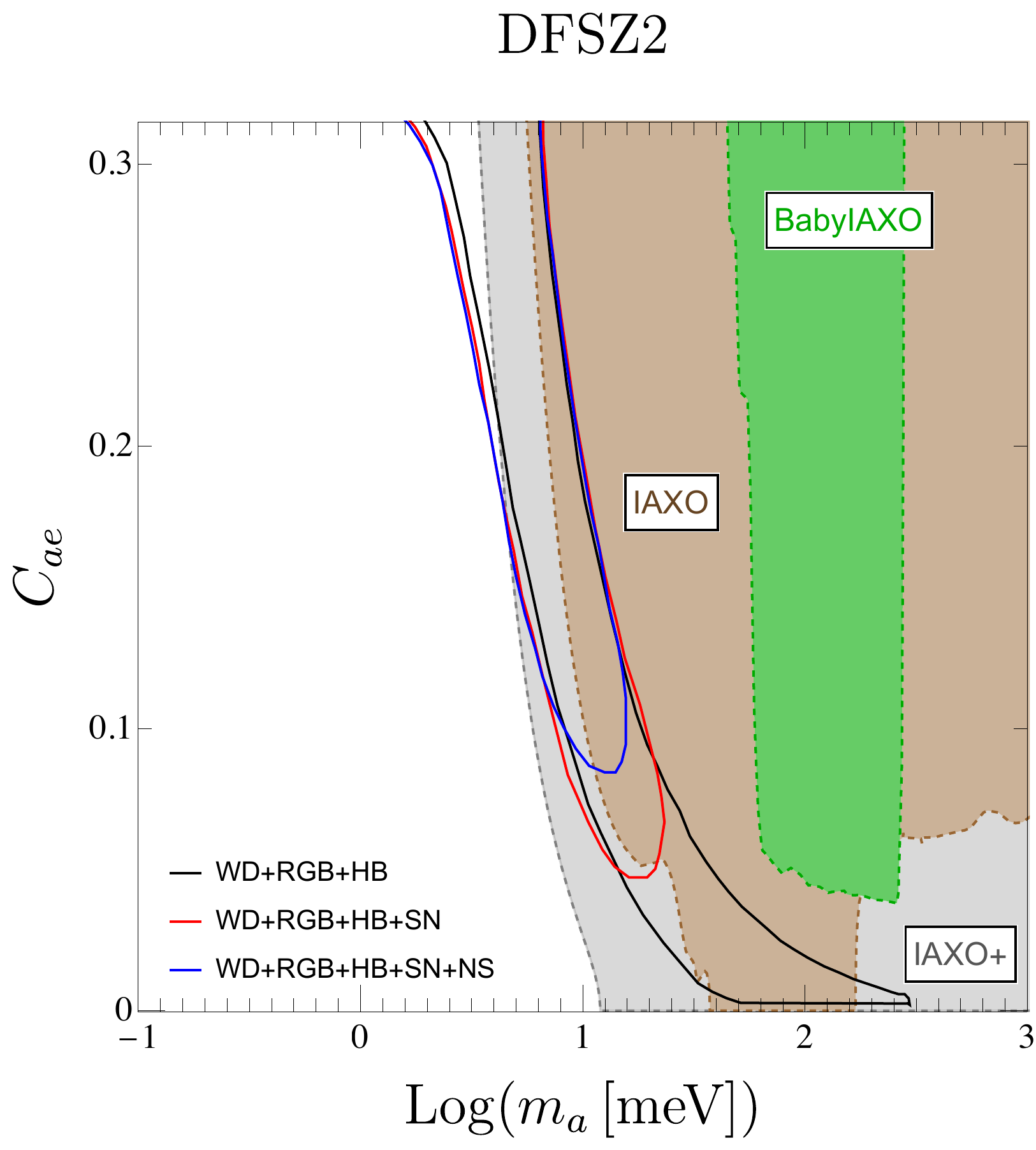}
\includegraphics[width=0.37\textwidth]{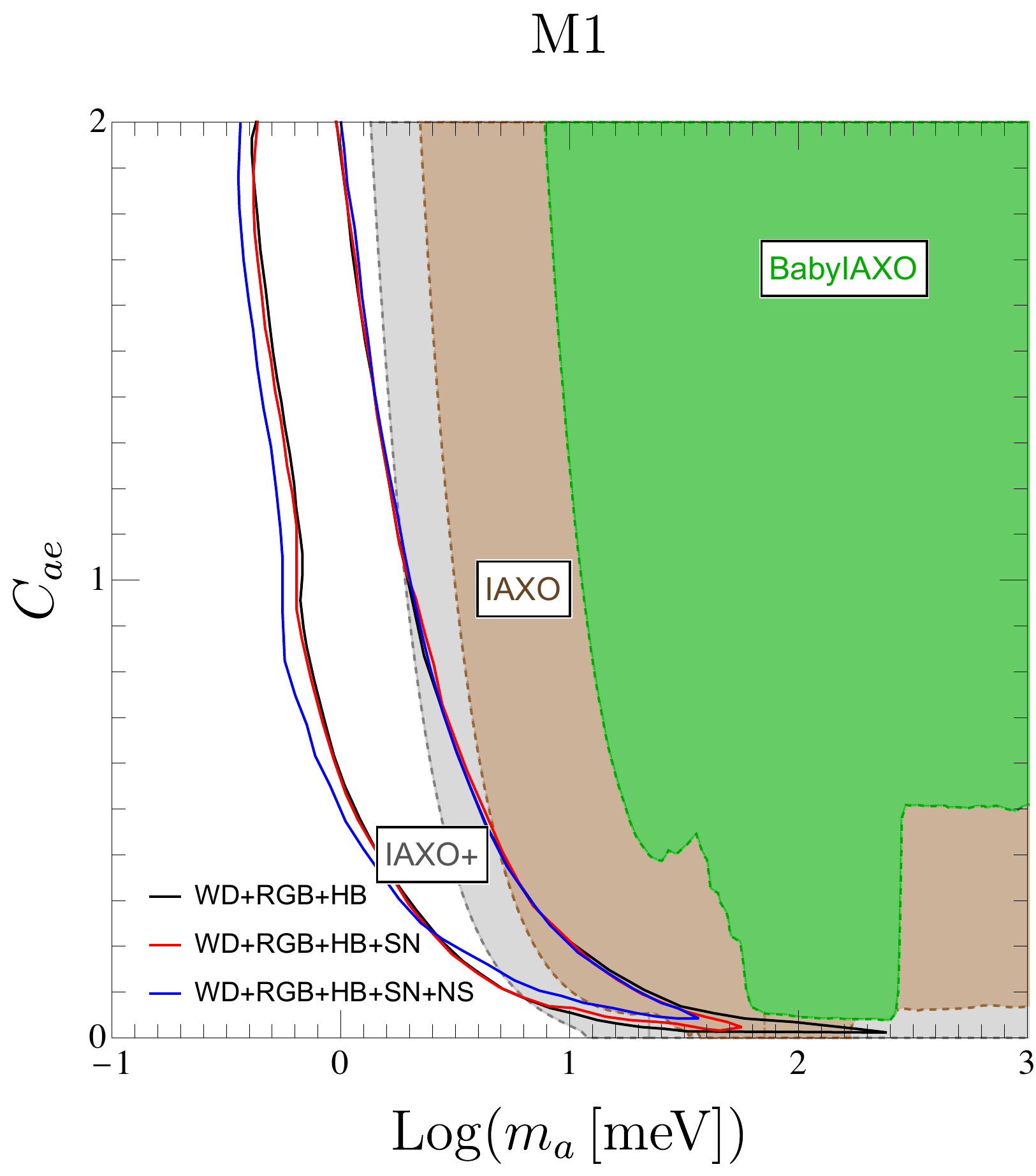}
\includegraphics[width=0.37\textwidth]{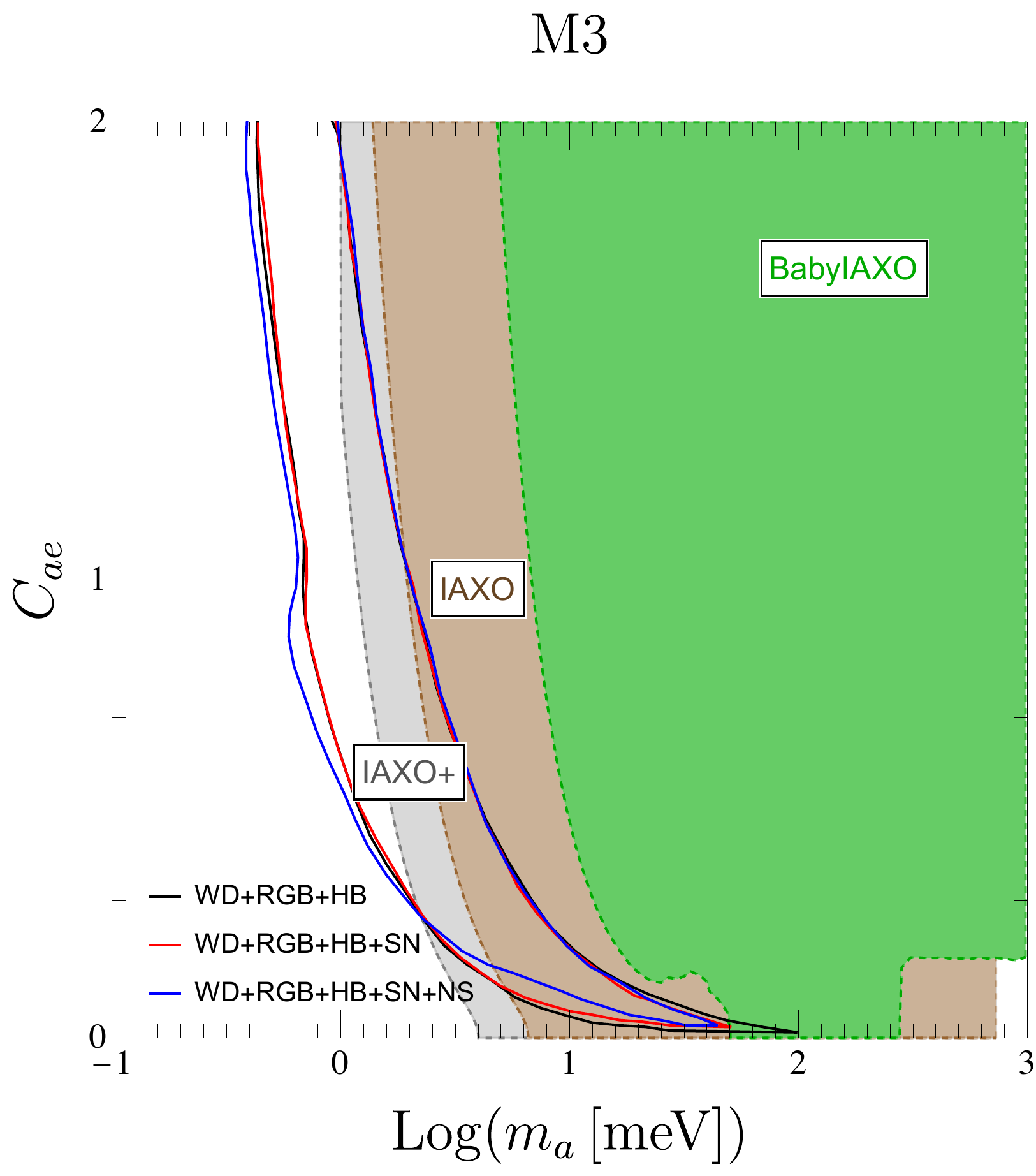}
\includegraphics[width=0.37\textwidth]{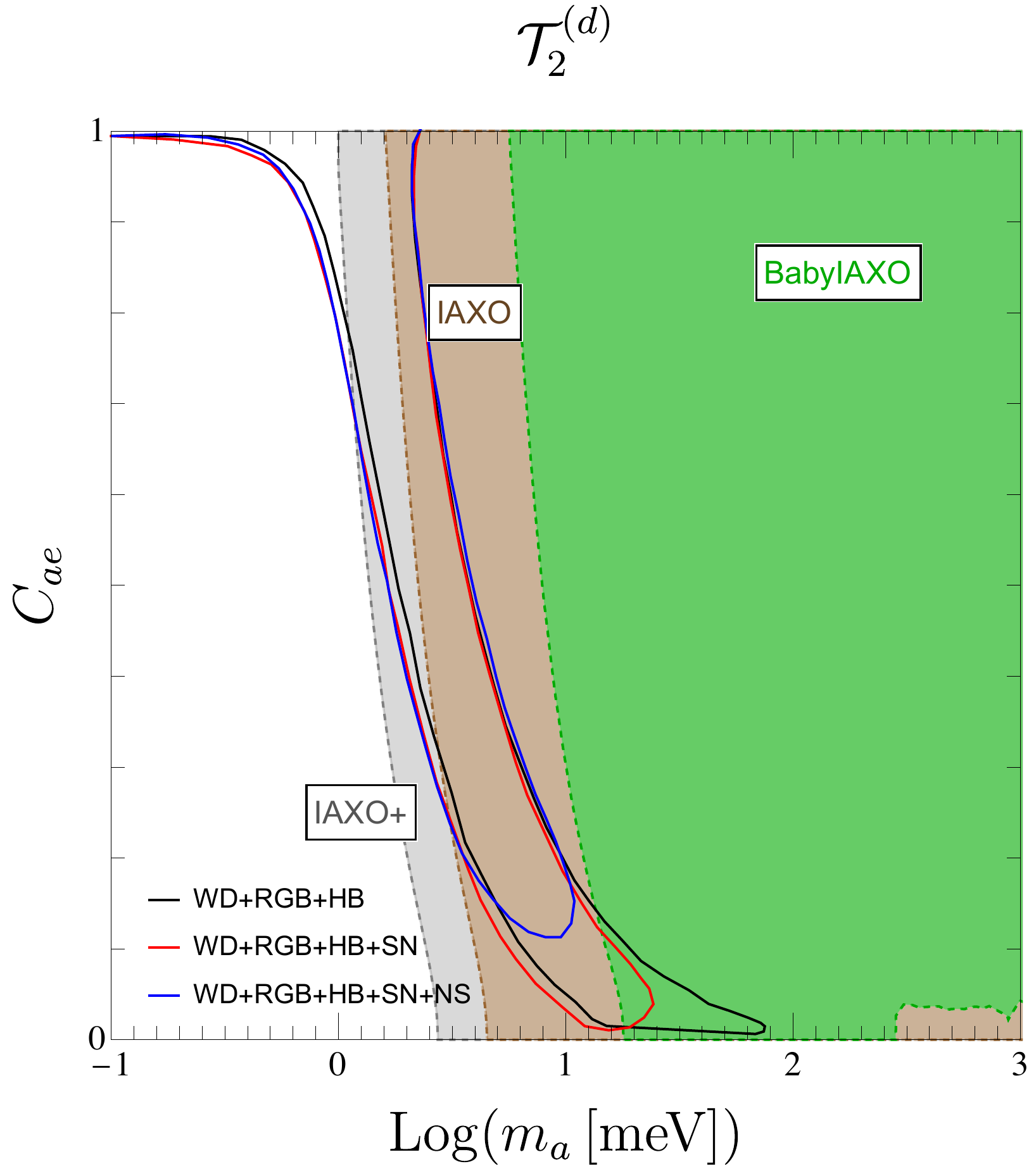}
\includegraphics[width=0.37\textwidth]{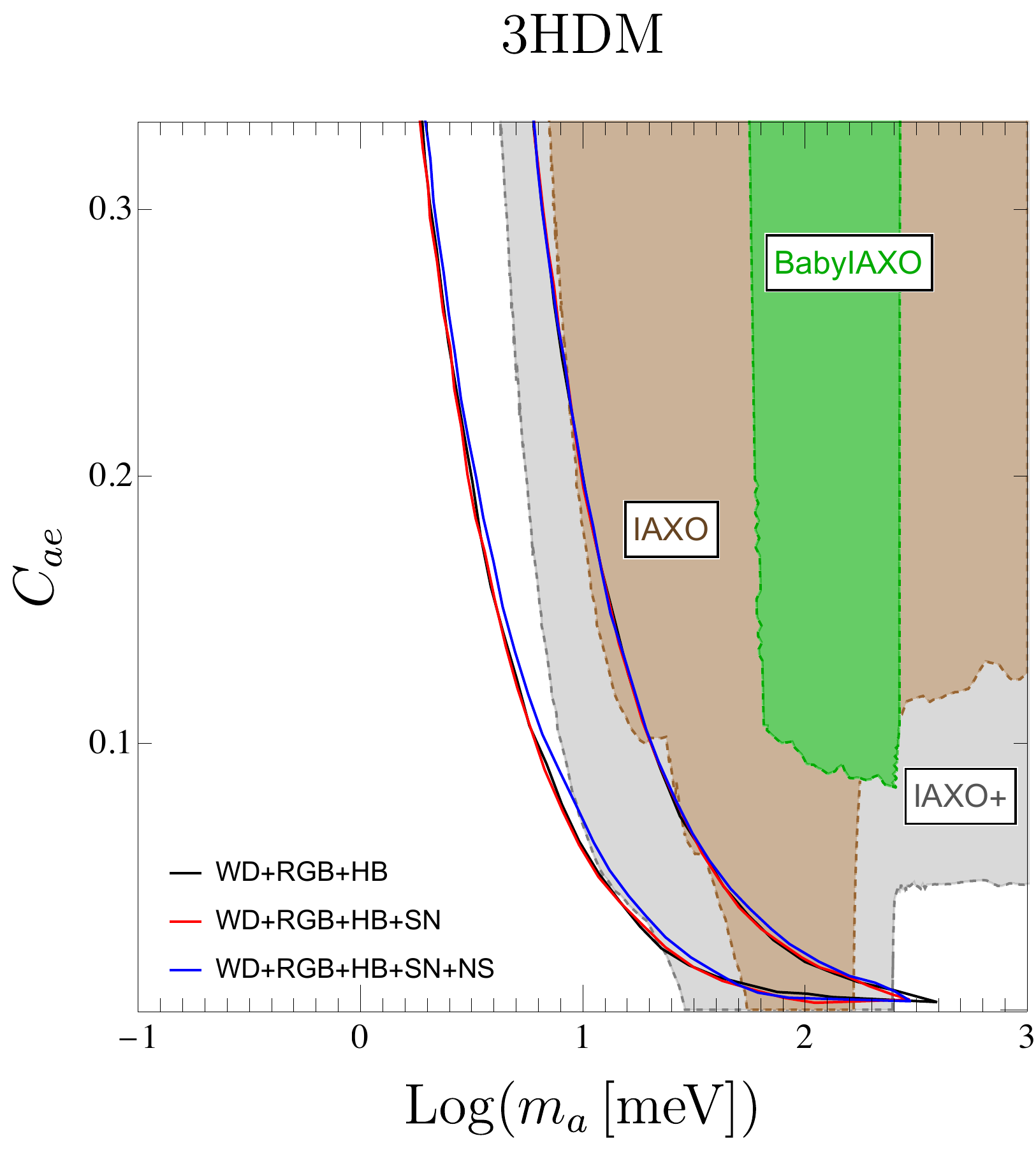}
\caption{Contours of the $2\sigma$ regions from global fits to stellar cooling in the $m_a$ vs. $C_{ae}$ plane for different models, following the colour scheme defined under Fig.~\ref{fig:gaegag_md} (continuous lines).
The colored regions show the helioscope sensitivities of BabyIAXO (in green), IAXO (in brown) and IAXO+ (in gray).}
\label{fig:helioscope_potential_ma}
\end{figure}
Finally, IAXO+  represents a more aggressive version of the IAXO helioscope, with a larger magnet and better optics.
The instrumental characteristics used in this work are extracted from Ref.~\cite{IAXO:2019mpb}.

The helioscope potential to detect the axion models discussed in the text is shown in the six 
panels of Fig.~\ref{fig:helioscope_potential_ma}.
In all cases, we show also the regions preferred by stellar evolution, 
making different assumptions for the astrophysical observables.
The areas corresponding to the analysis of WDs, RGB and the $R$ parameter (black contours) are the 
most solid, since the physics of these stars is relatively well known.
The regions drawn by including also SN 1987A and data from various NS, 
for which  the  physics of axion emission is still not completely understood, 
are shown with separate curves (red and blue contours respectively).

In all models, the preferred parameter region appears as a relatively narrow mass band, 
spanning from a few meV for large electron couplings, to a few 10 meV in the opposite limit. 
The strong dependence of the mass boundaries on the axion-electron coupling should be clear 
from our previous discussion:  as argued in Sec.~\ref{sec:modindepfit} (see, in particular, Fig.~\ref{fig:gaegag}) increasing the axion-electron coupling requires a smaller axion-photon 
coupling (and thus a smaller mass) to preserve the consistency with the observations. 
The axion-coupling with nucleons comes into the game  when the SN or the SN+NS observables are also included. In the case of nucleo-phobic models their role is obviously reduced. 
In general, however, they play a more important role at low values of the axion-electron coupling,
when the weight of  WD and RGB data is decreased. 

It is clear from the figures that the helioscopes may have a chance to detect the high mass end of the region favoured  by stellar evolution. 
Not surprisingly, the models most accessible to the helioscopes are those with a large coupling to photons, particularly $\mathcal{T}_2^{(u,d)}$ (cf.~Table \ref{tab:axionmodels}).
However, in general, the helioscope sensitivity increases (covering smaller masses) when the axion coupling to electrons increases, since this  guarantees a larger solar flux through the ABC 
production processes. This feature is discernible for all the models.

A rather interesting outcome of our analysis is the  potential
to explore the astrophysically interesting regions of the axion parameter space for some models, and in particular for $\mathcal{T}_2^{(u,d)}$, already with BabyIAXO. 
The upgraded versions, IAXO and especially IAXO+, will be clearly able to explore much larger sections of the parameter space. 

Besides helioscopes, other experiments may also probe these regions of the axion parameter space. 
Normally, however, such experiments rely on a set of additional assumptions, for example that axions are a substantial fraction of the local DM density.
In this case, preliminary studies~\cite{Marsh:2018dlj,Schutte-Engel:2021bqm} 
show that 
a new generation of haloscope detectors, based on Axionic Topological Antiferromagnets, might have the potential to explore 
the mass region of a few meV ($\lesssim 10\,$meV), 
possibly down to the axion photon couplings expected in some realistic axion models
(cf.~Fig.~23 in Ref.~\cite{Schutte-Engel:2021bqm}).
Another experiment with the potential to probe the axion mass up to several meV is ARIADNE~\cite{Arvanitaki:2014dfa}.
This is a long range force experiment, which relies on the axion coupling to 
nucleons. 
The experiment sensitivity, however, depends on the CP-violating axion-scalar coupling, 
whose phenomenologically allowed value spans several orders of magnitude~\cite{Moody:1984ba,Georgi:1986kr,Bertolini:2020hjc}.
Assuming the maximal experimentally allowed CP-violation beyond the SM, 
ARIADNE could potentially probe the DFSZ axion up to masses of several meV (cf.~also Fig.~2 in Ref.~\cite{Giannotti:2017hny}).

\section{Summary and conclusions}
\label{sec:concl}

In this paper we have revisited the global analysis of the stellar observations in relation 
to the axion couplings to the SM fields, providing an update with respect to 
Ref.~\cite{Giannotti:2017hny}  that includes the most recent observational results, 
and enlarging considerably the collection of axion models that are confronted with the data. 
The set of observations used in the present work is discussed in Sec.~\ref{sec:astroobs} while the statistical methodology is described in Sec.~\ref{sec:cooling fits}.
As regards the observational information,  the most significant update has been the inclusion of two recent analyses of the RGB bound on the axion-electron coupling, Refs.~\cite{Capozzi:2020cbu,Straniero:2020iyi}. 
These two studies have revised the previous bound and have refined the 
analysis taking advantage of the new  Gaia Data Release 2~\cite{chen2018}, which has significantly 
improved over the previous knowledge of the distance determinations. 
Furthermore, we have also included new analyses of the SN~\cite{Carenza:2019pxu} and NS~\cite{Beznogov:2018fda} bounds on the axion-nucleon couplings.

Our study confirms a preference for non-vanishing axion couplings to electrons and photons (cf.~Fig.~\ref{fig:gaegag}).
The hint is particularly strong for the axion-electron coupling, for which 
the best fit value lies away from zero with a significance of about $\sim 3\,\sigma$.
Our best fit values for the axion-electron and axion-photon couplings are 
$g_{ae} \simeq 1.2 \times 10^{-13}$,
 $g_{a\gamma} \simeq 1.8 \times 10^{-11}\,{\rm GeV^{-1}}$,
which are in good agreement with previous results,  
although, because of the new RGB analyses, the $g_{ae}$ best fit point is slightly shifted 
to a lower value.

To investigate the theoretical implications of these hints, we have studied 
a set of well motivated QCD axion models whose relevant features 
have been summarised  in Sec.~\ref{sec:models}. 
On top of the two 
benchmark DFSZ axion models, 
we have considered a general class 
of non-universal DFSZ-like  models, 
featuring generation-dependent PQ charges. 
The logic behind the models' selection was to have in first place a sizeable 
$C_{ae} / C_{a\gamma}$ ratio, a favorable condition that was  
highlighted by the model-independent fit in \fig{fig:gaegag}, 
and next to allow for the possibility of suppressing sufficiently the axion-nucleon 
couplings.  This latter condition allows to circumvent to some extent 
the strong bounds from SN 1987A and from NS.
The results of our analysis for the different models are presented in 
Fig.~\ref{fig:helioscope_potential_ma}.
The qualitative feature is the same for all models:
for large electron couplings, the axion photon coupling is pushed towards  relatively  small values (cf.~Fig~\ref{fig:gaegag}), and correspondingly  the preferred region for the axion mass 
is also driven to relatively small values $m_a\sim 1\,$meV. 
In the case of small axion-electron coupling, on the other hand, there is a preference for a more sizeable axion-photon coupling and thus for a larger axion mass, up to 100 meV or so, depending on the model.
In all cases, the inclusion of constraints on the axion-nucleon couplings from SN 1987A or from 
the SN and the NS pushes the mass to lower values, since these observations prefer a vanishing axion-nucleon coupling (and hence a large value of the axion decay constant $f_a$ which 
suppresses the axion mass). In the case of nucleo-phobic models however, axion-nucleon 
decoupling is assisted by a conspiracy between the values of the PQ charges of the quarks, 
and accordingly the mass suppression effect is considerably less important. 

An important outcome of our study is that star evolution observations 
indicate a clear preference for an axion mass in the meV region,
a range that is at least partially accessible to the next generation of axion helioscopes.
The discovery  potential of helioscopes is quite sensitive to the axion-photon coupling, which is largest in the  $\mathcal{T}_2^{(u,d)}$ models (the complete list of axion-photon couplings 
for all non-universal DFSZ axion models with two Higgs doublets is given in Tab.~\ref{tab:EoNgagamma}).
It is quite remarkable that already BabyIAXO will be able to probe some of these models, and that IAXO and IAXO+ will have a chance to cover the hinted region entirely. 

It should be also stressed that a direct exploration of this region of parameter space with 
dedicated axion experiments will have an impact not only for axion physics but also for astrophysics.
Although the  stellar hints individually do not have a large significance, they all show a preference 
for an increased  efficiency in star energy loss. Assuming no new physics is at play, 
this would signal a systematic problem in our understanding of stellar cooling, a possibility that 
would gain strength if some preferred particle physics explanations could be ruled out. 
Certainly, the improvements in the astrophysical instrumentation expected in the next decade or 
so will have a strong impact for clarifying some of these issues, and it might eventually 
strengthen the case for new energy loss channels. 
 In such scenario, a new physics solution, perhaps in the form of axions, would be a most exciting result. In any case, whether this problem requires new physics, or is just a matter of 
  understanding better the details of star evolution, or is merely an instrument 
  calibration issue, will be  clarified in the (hopefully) near future.
In the meanwhile, we keep being fascinated by the synergy between astrophysics 
and particle physics, and by the possibility that the first evidences for the need 
of new particle physics could eventually come from new careful observations of the sky.

\begin{small}

\section*{Acknowledgments} 
We thank 
Sebastian Hoof, 
Igor Irastorza 
and Javier Redondo for useful discussions.  
M.F., M.G.~and F.M.~acknowledge the INFN Laboratori Nazionali di Frascati for hospitality and partial financial support during the completion of this project.
The work of L.D.L.~is supported by the Marie Sk\l{}odowska-Curie 
Individual Fellowship grant AXIONRUSH (GA 840791) and the Deutsche Forschungsgemeinschaft under Germany's Excellence Strategy 
- EXC 2121 Quantum Universe - 390833306. 
L.D.L.~is also supported by the 
European Union's Horizon 2020 research and innovation programme under the Marie Sk\l{}odowska-Curie grant agreement No 860881-HIDDEN.
The work of M.G.~is partially supported by a grant provided by the Fulbright U.S.~Scholar Program and by a grant from the Fundación Bancaria Ibercaja y Fundación CAI.
M.G.~thanks the Departamento de Física Teórica and the Centro de Astropartículas y Física de Altas Energías (CAPA) of the Universidad de Zaragoza for hospitality during the completion of this work. 
F.M.~acknowledges financial support from the State Agency for Research of the Spanish Ministry of Science and Innovation
through the “Unit of Excellence Mar\'ia de Maeztu 2020-2023” award to the Institute of Cosmos
Sciences (CEX2019-000918-M) and from PID2019-105614GB-C21 and 2017-SGR-929 grants.
The work of M.F.~is supported  by project C3b of the DFG-funded Collaborative Research Center TRR 257, ``Particle Physics Phenomenology after the Higgs Discovery''.  E.N.~is supported by the INFN {\it Iniziativa Specifica} 
Theoretical Astroparticle Physics (TAsP-LNF).

\end{small}

\appendix

\section{\texorpdfstring{$E/N$}{E/N} factors in non-universal DFSZ models}
\label{sec:generalDFSZ}

In this Appendix we work out the 
discrete set of $E/N$ values that 
can be obtained
in the most general non-universal DFSZ
axion model with two Higgs doublets. 

Requiring that the determinants of the mass matrices of the quarks does not vanish
yields a relation between the QCD anomaly and the charges $\mX_{ij}$
of the Higgs doublets. Denote as  $H_{il}$, with charge $\mX_{il}$,  the Higgs coupled to the ($il$) entry of the mass matrix.
Let us take $\det M = \epsilon_{ijk}\epsilon_{lmn} \,M_{il}\, M_{jm} \, M_{kn} \neq 0$. This 
means that at least one term in the sum must be different from zero. 
Take for example  $M_{il} M_{jm} M_{kn} \neq 0$.
In terms of charges this means 
(e.g.~for up-type quarks) 
$-\mX_{q_i} +\mX_{u_l} + \mX_{i l}=  0$
and other two similar relations for ($jm$) and ($kn$).
Since no pair of indices can be equal in the two triplets  $(i,j,k)$ and $(l,m,n)$,  
from the non-vanishing product of the three  matrix elements  it follows
\beq
\label{eq:up}
\sum_{i=1}^3 \mX_{q_i} - \sum_{i=1}^3 \mX_{u_i} = \mX_{il} +\mX_{jm} + \mX_{kn} \quad \to \quad
3 \mX_1, \ \  \mathrm{or} \ \ 2  \mX_1 + \mX_2\,, 
\eeq
where in the last step we assumed only two Higgs doublets, and we have defined as  $H_1$ the doublet that occurs more 
times in the product of the three mass matrix entries   of the non-vanishing term of the determinant.  
Doing the same for the down-type quarks and leptons is straightforward. However,  now 
we have to keep the distinct  possibilities for $\mX_1 \leftrightarrow \mX_2$. So in the two Higgs doublet case we get  
\beq
\label{eq:down}
\sum_{i=1}^3 \mX_{q_i} - \sum_{i=1}^3 \mX_{d_i} = - 3 \mX_1\,, \ \ \mathrm{or} \ \ - 2  \mX_1 - \mX_2\,, \ \ \mathrm{or} \ \ -  \mX_1 - 2 \mX_2\,, \ \ \mathrm{or} \ \ 
- 3 \mX_2\,.
\eeq
Considering together Eq.~\eqref{eq:up} and Eq.~\eqref{eq:down}
we see that we have $2 \times 4$ different possibility for the  QCD anomaly (some will give the same result). 
For the numerator of the second term of our $E/N$ expression, that reads 
$\sum_i \mX_{q_i} - \sum_i \mX_{u_i} + \sum_i \mX_{\ell_i} - \sum_i \mX_{e_i}$, 
one gets the same $2\times 4$ possibilities,  since \eq{eq:down} holds also for the leptons.
Then, by combining in all the ways  the  possible numerators and denominators, one gets 
\begin{align}
\dfrac{E}{N} &= 
 2 N_c\, q^2_d + 2q^2_e\, \dfrac{\sum_i \(\X_{u_i} - \X_{e_i}\)}{\sum_j\(\X_{u_i} - \X_{d_j}\)} 
 = 
\dfrac{2}{3}+ 2 (\pm 2, \pm 1, \pm \dfrac{1}{2}, 0, \dfrac{1}{3}, \dfrac{2}{3}, \dfrac{3}{2}, 3) \, ,
\end{align}
corresponding to the results 
displayed in Table~\ref{tab:EoNgagamma}. 
\begin{table}[!htp]
\begin{center}
\begin{tabular}{|c|c|c|}
\hline
$E/N$ & $|C_{a\gamma}|$ & Models\\
\hline
$-10/3$  &  $5.25$  & $\mathcal{T}_2^{(u)}$\\
$20/3$  &  $4.75$    &$\mathcal{T}_2^{(d)}$ \\
$-4/3$ & $3.25$  & M3 \\
$14/3$ & $2.75$ & M4\\
$-1/3$ & $2.25$ & -\\
$11/3$ & $1.75$ & -\\
$2/3$ & $1.25$ & M1 \\
$8/3$ & $0.75$ & M2 \\
$4/3$ & $0.59$ & -\\
$5/3$ & $0.25$ & -\\
$2$ & $0.08$ & -\\
\hline
\end{tabular}
\end{center}
\caption{List of all the possible values of $E/N$ in DFSZ-like models with two Higgs doublets, ordered by larger contribution to $g_{a\gamma}$. The canonical $E/N$ values for DFSZ1 and DFSZ2 coincide with those for M2 and M1, respectively.}
\label{tab:EoNgagamma}
\end{table}

\bibliographystyle{JHEP.bst}
\bibliography{bibliography}

\end{document}